\documentclass[11pt]{article}

\include{setup}

\begin{document}

\thispagestyle{empty}
\def\thefootnote{\fnsymbol{footnote}}
\setcounter{footnote}{1}
\null
\draftdate\hfill FR-PHENO-2012-034\\
\strut\hfill SFB/CPP-11-62\\
\strut\hfill TTK-12-47\\
\strut\hfill TTP11-23\\
\strut\hfill LPN11-62\\
\vskip 0cm
\vfill
\begin{center}
  {\Large \boldmath{\bf Electroweak corrections to monojet production
      \\
      at the Tevatron and the LHC}
\par} \vskip 2.5em
{\large
{\sc Ansgar Denner$^{1}$, Stefan Dittmaier$^{2}$, 
     Tobias Kasprzik$^{3}$, Alexander M\"uck$^{4}$
}\\[2ex]
{\normalsize \it $^1$Universit\"at W\"urzburg, 
Institut f\"ur Theoretische Physik und Astrophysik,\\ 
D-97074 W\"urzburg, Germany}
\\[1ex]
{\normalsize \it 
$^2$Albert-Ludwigs-Universit\"at Freiburg, 
Physikalisches Institut, \\
D-79104 Freiburg, Germany
}\\[1ex]
{\normalsize \it 
$^3$Karlsruhe Institute of Technology (KIT), 
Institut f\"ur Theoretische Teilchenphysik,\\
D-76128 Karlsruhe, Germany
}\\[1ex]
{\normalsize \it 
$^4$RWTH Aachen University, 
Institut f\"ur Theoretische Teilchenphysik und Kosmologie,\\
D-52056 Aachen, Germany
}\\[2ex]
}
\par \vskip 1em
\end{center}\par
\vskip .0cm \vfill {\bf Abstract:} \par
Single-jet production with missing transverse momentum is one of
the most promising discovery channels for new physics at the LHC. 
In the Standard Model, $\PZ + \mathrm{jet}$ production with a
\PZ-boson decay into neutrinos leads to this monojet signature.  To
improve the corresponding Standard Model predictions, we present the
calculation of the full next-to-leading-order (NLO) electroweak corrections 
and a recalculation of the NLO QCD corrections to
monojet production at the Tevatron and the LHC. We discuss the
phenomenological impact on the total cross sections as well as on
relevant differential distributions.

\par\vskip 1cm
\noindent
November 2012    
\par
\null
\setcounter{page}{0}
\clearpage
\def\thefootnote{\arabic{footnote}}
\setcounter{footnote}{0}

\section{Introduction}

Many models for Beyond Standard Model (BSM) physics predict new
particles which are not visible for a typical detector in a collider
experiment. When such an invisible particle is produced and recoils
against a QCD jet, the detector will measure a monojet event, i.e.\ an
event which only shows a single jet with potentially large transverse
momentum and an equal amount of missing transverse momentum.  For
instance, such monojet signatures are predicted in extra-dimension
models, (see, e.g., \citere{ArkaniHamed:1998rs}), where single
(undetectable) gravitons are produced together with hard jets (see
Refs.~\cite{Karg:2009xk, Karg:2010zz} for a next-to-leading-order (NLO)
QCD analysis). Moreover, certain models including stable
unparticles~\cite{Georgi:2007ek}, as well as the decay of
mini-black-holes, which might be produced at the LHC according to
large-extra-dimension scenarios (see, e.g., Ref.~\cite{Betz:2006ds} and
references therein), may also give rise to single-jet
events. Accordingly, it has been investigated how to discriminate the
different models via their predicted phenomenological signatures at the
LHC~\cite{Rizzo:2008fp}.  Recently, first experimental analyses of LHC
monojet data have been accomplished~\cite{Aad:2011xw,Chatrchyan:2011nd}
to constrain the ADD model of Large Extra
Dimensions~\cite{ArkaniHamed:1998rs} and to search for dark matter,
finding agreement with the Standard Model (SM) predictions.

In the SM, monojets are produced in $\PZ + \mathrm{jet}$
events if the \PZ\ boson is decaying into neutrinos. 
Hence, any BSM search in the monojet channel relies on the precise
predictions for this SM process.
Moreover, monojet events from $\PZ + \mathrm{jet}$
production have a relatively large cross section and can be used to
study the measurement of missing transverse momentum $\ptmiss$ in
general.  A proper understanding of missing transverse momentum
$\ptmiss$ is relevant in an even wider class of BSM searches, in
particular for searches in models containing a TeV-scale dark-matter
candidate. As stated in \citeres{Vacavant:2001sd,Benucci:2010ax}, for
low values of $\ptmiss$, uncertainties
within the SM mostly arise from uncertainties in the jet-energy
measurement. At high $\ptmiss$, however, the SM background is indeed
dominated by on-shell $\PZ + \mathrm{jet}$ production with the
subsequent decay $\PZ \to \Pnl\Pnlb$. Of course, there are further
sources for monojet events in the SM, for instance \PW\ + jet
production, where the charged lepton from the \PW-boson decay is not
reconstructed, which are not considered in this work.

BSM searches will probe monojets with larger and larger transverse momenta
with increasing centre-of-mass (CM) energy and luminosity at the
LHC. Since electroweak (EW) radiative corrections grow with energy and are
known to reach tens of percent at the TeV scale, they
become more and more relevant and cannot be neglected compared to the
usually included corrections from strong interaction.
For instance, in this paper we show that EW corrections
reduce the SM monojet cross section by about $15\%$ at a missing transverse
momentum of only $500\GeV$. Based on these motivations, in this
work, we investigate in detail $\PZ + \mathrm{jet}$ production
including the $\PZ$-boson decay into neutrinos, i.e.\
\begin{equation}
\mathrm{pp/p\bar{p}} \to \PZ + \mathrm{jet} +X \to \Pnl\Pnlb + \mathrm{jet}
+ X \, .
\end{equation}
In particular, we provide the EW and QCD corrections at NLO
not relying on any on-shell approximation for the intermediate \PZ\ boson.

From the QCD point of view, the description of 
\PZ\ + jet production with leptonically decaying Z~bosons
hardly depends on the specific final-state leptons in any order
in perturbation theory, since the leptonic decay products are
insensitive to strong contributions. 
For on-shell Z~bosons, which appear, e.g., in the treatment via
the narrow-Z-width approximation, this means that the relative
QCD corrections are indeed identical for 
$\PZ(\to\Plm\Plp/\Pnl\Pnlb)+ \mathrm{jet}$.
Including Z-boson off-shell effects, however, leads to differences
between the two leptonic final states because of the different 
spin correlations due to the different chiral couplings of $\Pl$
and $\Pnl$ and, more importantly the existence of $\gamma^*$ exchange
for charged leptons.
The NLO corrections to \PZ\ + jet production are known for a long
time~\cite{Giele:1993dj,Campbell:2002tg,vanderBij:1988ac} and were
matched with parton showers~\cite{Alioli:2010qp}. A resummation of large
logarithms at the next-to-leading-logarithmic
level was performed in Ref.~\cite{Becher:2011fc} for vector-boson
production at large transverse momentum.

Considering EW corrections, the situation is more involved,
because of non-factorizable corrections due to interactions of
initial-state partons and final-state leptons. In a first step, the
purely weak one-loop corrections to $\PZ + \mathrm{jet}$ production in
the SM were investigated in the on-shell
approximation~\cite{Kuhn:2004em,Kuhn:2005az}, 
i.e.\ with a stable external \PZ~boson,
and photonic corrections were ignored. For \PZ\ bosons at large transverse
momentum, requiring a large CM energy, using on-shell \PZ\
bosons is a good approximation since the EW corrections are dominated by
large universal Sudakov logarithms~\cite{Ciafaloni:1998xg}. In
\citere{Kuhn:2004em} the leading corrections up to the next-to-leading
logarithms at the one- and two-loop level were calculated.  Later
the full NLO weak corrections were added~\cite{Kuhn:2005az}.  
Although all off-shell effects are neglected, one would still expect the
$\ptmiss$ distribution in the monojet scenario to be described well by
the on-shell computation, at least at high energies. However, aiming at 
a precision at the percent level,
photonic corrections have to be taken into account, as well as
off-shell effects due to the decay of the virtual boson. These issues
are resolved in this work for monojet production at the LHC and the
Tevatron.

There are other uncertainties at the level of several percent affecting
the cross section for monojet production that are not addressed in this
work. These result, in particular, from missing higher-order QCD
corrections, from the modelling of parton showers and fragmentation,
from parton distribution functions (PDFs), and from the limited
knowledge of the jet energy scale and resolution.  Estimates on these
uncertainties can be found in experimental publications on $\PZ +
\mathrm{jet}$ production~\cite{Aad:2011qv, Chatrchyan:2011wt}, in theory
papers \cite{Ita:2011wn}, and in discussions of PDF uncertainties
\cite{Watt:2011kp}. These uncertainties will be further reduced by
improved theoretical predictions and future experimental analyses. In
the search for new physics in monojet production, the SM background
cross section is typically estimated from experimental data for
$\Pl^+\Pl^- + \mathrm{jet}$ events. In the analysis of
Ref.~\cite{Aad:2011xw}, based on an integrated luminosity of 33
pb$^{-1}$, the systematic uncertainties on the SM monojet event rates
are at the level of 10\% and result dominantly from limited statistics
in the control region. With improved statistics the EW corrections may
become even more important, in particular since many uncertainties
cancel in the cross-section ratio of $\Pl^+\Pl^-+\mathrm{jet}$ and
$\nu_l\bar{\nu}_l+\mathrm{jet}$ production.

Besides the new evaluation of the complete EW corrections for the
off-shell case we have also recalculated the well-known NLO QCD
corrections, supporting a phase-space dependent renormalization and
factorization scale.  Our results are implemented in a flexible Monte
Carlo code that allows for the computation of total cross sections as
well as differential distributions.  Our implementation is completely
generic in the sense that there are no restrictions for the
event-selection criteria to be applied. All off-shell effects are
included, and the finite \PZ-boson width is consistently taken care of
using the complex-mass scheme~\cite{Denner:2005fg}.  Since the
computation for the $\Pnl\Pnlb + \mathrm{jet}$ final state is closely
related to the final state with two charged leptons, the presentation in
this paper follows~\citere{Denner:2011vu}, where the calculation of the
EW corrections to $\Plm\Plp + \mathrm{jet}$ production at hadron
colliders was presented.

This paper is organized as follows. In \refse{se:details}, we briefly
describe our calculation and discuss the relevant
theoretical concepts. In
\refse{se:numres}, we specify the numerical input as well as the details
of our event selection.  Numerical results are given for monojet
production both at the LHC and at the Tevatron. We present inclusive
cross sections for specified sets of cuts and distributions for
the jet transverse momentum and rapidity, as well as the missing
transverse momentum, and discuss the impact of the EW
contributions. Section~\ref{se:EWQCD} discusses a proper combination
of the EW corrections with existing
higher-order QCD predictions. We conclude in~\refse{se:concl}.

\section{Details of the calculation}
\label{se:details} 

\begin{sloppypar}
  This calculation closely follows the one for charged dilepton + jet
  production presented in~\citere{Denner:2011vu}.  Here we only briefly
  summarize the essential ingredients of our calculation and point to
  the differences with respect to \citere{Denner:2011vu}. More details
  can be found in Section~2 of that paper.
\end{sloppypar}

\subsection{General setup}
\label{se:setup} 

In the SM, at leading order (LO) in perturbation theory 
the transverse momentum of a single isolated jet is
balanced by a \PZ~boson, which decays into two undetected neutrinos. 
At hadron colliders, the 
partonic channels 
\begin{eqnarray}
\label{eq:proc1}
& \Pq_i \, \, \Pqbar_i & \to \PZ\, \Pg      \to \Pnl \Pnlb \, \Pg \, ,\\
\label{eq:proc2}
& \Pq_i \, \, \Pg      & \to \PZ\, \Pq_i    \to \Pnl \Pnlb \, \Pq_i \, ,\\
\label{eq:proc3}
& \Pqbar_i \, \, \Pg   & \to \PZ\, \Pqbar_i \to \Pnl \Pnlb \, \Pqbar_i \, 
\end{eqnarray}
have to be taken into account, where $\Pq_i$ denotes any light quark, i.e.\
$\Pq_i=\Pu,\Pd,\Pc,\Ps,\Pb$. 
Note that we always imply the summation over neutrino
flavours in the final state, leading to a trivial factor of three for 
all cross sections compared to the cross sections for a single
neutrino flavour. The tree-level Feynman diagrams for process
\refeq{eq:proc1} are shown in \reffi{fi:born_qqZg}. 
The intermediate \PZ-boson resonance is treated in a gauge-invariant way using the
complex-mass scheme~\cite{Denner:2005fg}.
\bfi
\begin{center}
\unitlength=3bp%

\begin{center}
\begin{small}
\begin{feynartspicture}(80,30)(2,1)

\FADiagram{}
\FAProp(0.,15.)(10.,5.5)(0.,){/Straight}{1}
\FALabel(3.19219,13.2012)[bl]{$q$}
\FAProp(0.,5.)(10.,13.)(0.,){/Straight}{-1}
\FALabel(3.17617,6.28478)[tl]{$q$}
\FAProp(20.,17.)(16.,13.5)(0.,){/Straight}{-1}
\FALabel(17.4593,15.9365)[br]{$\nu_l$}
\FAProp(20.,10.)(16.,13.5)(0.,){/Straight}{1}
\FALabel(17.4593,11.0635)[tr]{$\nu_l$}
\FAProp(20.,3.)(10.,5.5)(0.,){/Cycles}{0}
\FALabel(14.4543,2.54718)[t]{$\Pg$}
\FAProp(10.,5.5)(10.,13.)(0.,){/Straight}{1}
\FALabel(11.07,9.25)[l]{$q$}
\FAProp(10.,13.)(16.,13.5)(0.,){/Sine}{0}
\FALabel(12.8713,14.3146)[b]{$\PZ$}
\FAVert(10.,5.5){0}
\FAVert(10.,13.){0}
\FAVert(16.,13.5){0}

\FADiagram{}
\FAProp(0.,15.)(10.,13.)(0.,){/Straight}{1}
\FALabel(5.30398,15.0399)[b]{$q$}
\FAProp(0.,5.)(10.,5.5)(0.,){/Straight}{-1}
\FALabel(5.0774,4.18193)[t]{$q$}
\FAProp(20.,17.)(15.5,13.5)(0.,){/Straight}{-1}
\FALabel(17.2784,15.9935)[br]{$\nu_l$}
\FAProp(20.,10.)(15.5,13.5)(0.,){/Straight}{1}
\FALabel(18.2216,12.4935)[bl]{$\nu_l$}
\FAProp(20.,3.)(10.,5.5)(0.,){/Cycles}{0}
\FALabel(15.3759,5.27372)[b]{$\Pg$}
\FAProp(10.,13.)(10.,5.5)(0.,){/Straight}{1}
\FALabel(8.93,9.25)[r]{$q$}
\FAProp(10.,13.)(15.5,13.5)(0.,){/Sine}{0}
\FALabel(12.8903,12.1864)[t]{$\PZ$}
\FAVert(10.,13.){0}
\FAVert(10.,5.5){0}
\FAVert(15.5,13.5){0}

\end{feynartspicture}
\end{small}
\end{center}

\vspace*{-1.6em}
\end{center}
\mycaption{\label{fi:born_qqZg} Feynman diagrams for the LO process \refeq{eq:proc1}.}
\efi
Since we consider $\PZ+\mathrm{jet}$ production with a subsequent
\PZ-boson decay into neutrinos at NLO accuracy with respect to EW
corrections, i.e.\ at the order $\mathcal{O}(\alpha^3
\alpha_{\mathrm{s}})$, we also include the photon-induced processes,
\begin{eqnarray}
\label{eq:proc4}
& \Pq_i \, \, \ga      & \to \PZ\, \Pq_i    \to \Pnl \Pnlb \, \Pq_i \, ,\\
\label{eq:proc5}
& \Pqbar_i \, \, \ga   & \to \PZ\, \Pqbar_i \to \Pnl \Pnlb \, \Pqbar_i \, ,
\end{eqnarray}
which contribute at the order $\mathcal{O}(\alpha^3)$ and may thus be
relevant at the level of a few percent.
\bfi
\begin{center}
\unitlength=3bp%

\begin{center}
\begin{small}
\begin{feynartspicture}(80,30)(2,1)

\FADiagram{}
\FAProp(0.,15.)(5.5,10.)(0.,){/Straight}{1}
\FALabel(2.18736,11.8331)[tr]{$q$}
\FAProp(0.,5.)(5.5,10.)(0.,){/Sine}{0}
\FALabel(3.31264,6.83309)[tl]{$\gamma$}
\FAProp(20.,17.)(15.5,13.5)(0.,){/Straight}{-1}
\FALabel(17.2784,15.9935)[br]{$\nu_l$}
\FAProp(20.,10.)(15.5,13.5)(0.,){/Straight}{1}
\FALabel(18.2216,12.4935)[bl]{$\nu_l$}
\FAProp(20.,3.)(12.,10.)(0.,){/Straight}{-1}
\FALabel(15.4593,5.81351)[tr]{$q$}
\FAProp(5.5,10.)(12.,10.)(0.,){/Straight}{1}
\FALabel(8.75,8.93)[t]{$q$}
\FAProp(15.5,13.5)(12.,10.)(0.,){/Sine}{0}
\FALabel(13.134,12.366)[br]{$\PZ$}
\FAVert(5.5,10.){0}
\FAVert(15.5,13.5){0}
\FAVert(12.,10.){0}

\FADiagram{}
\FAProp(0.,15.)(10.,13.)(0.,){/Straight}{1}
\FALabel(5.30398,15.0399)[b]{$q$}
\FAProp(0.,5.)(10.,5.5)(0.,){/Sine}{0}
\FALabel(5.0774,4.18193)[t]{$\gamma$}
\FAProp(20.,17.)(15.5,13.5)(0.,){/Straight}{-1}
\FALabel(17.2784,15.9935)[br]{$\nu_l$}
\FAProp(20.,10.)(15.5,13.5)(0.,){/Straight}{1}
\FALabel(18.2216,12.4935)[bl]{$\nu_l$}
\FAProp(20.,3.)(10.,5.5)(0.,){/Straight}{-1}
\FALabel(15.3759,5.27372)[b]{$q$}
\FAProp(10.,13.)(10.,5.5)(0.,){/Straight}{1}
\FALabel(8.93,9.25)[r]{$q$}
\FAProp(10.,13.)(15.5,13.5)(0.,){/Sine}{0}
\FALabel(12.8903,12.1864)[t]{$\PZ$}
\FAVert(10.,13.){0}
\FAVert(10.,5.5){0}
\FAVert(15.5,13.5){0}

\end{feynartspicture}
\end{small}
\end{center}

\vspace*{-1.6em}
\end{center}
\mycaption{\label{fi:born_qgaZq} Feynman diagrams for the
  photon-induced process \refeq{eq:proc4}.}  \efi The tree-level
Feynman diagrams for process \refeq{eq:proc4} can be found in
\reffi{fi:born_qgaZq}.  We use the MRSTQED2004 
set~\cite{Martin:2004dh} of PDFs
to estimate the photon content of the proton
but employ modern PDF sets for all partonic channels without
initial-state photons.

The Feynman diagrams and amplitudes are generated with the {\sc
  FeynArts} package~\cite{Kublbeck:1990xc} and further processed with
{\sc Pole}~\cite{Accomando:2005ra} and {\sc
  FormCalc}~\cite{Hahn:1998yk}, or alternatively with independent
in-house {\sl Mathematica} routines.  Hence, all parts of the
calculation are again performed in two independent ways with different
tools.

\subsection{Virtual corrections}
\label{se:virt}

Virtual one-loop QCD and EW corrections are calculated for the
partonic processes \refeq{eq:proc1}--\refeq{eq:proc3}.  
We neglect the NLO QCD corrections to the photon-induced processes
\refeq{eq:proc4} and \refeq{eq:proc5}, which are expected to be as tiny
as for $\PW+\mathrm{jet}$ production~\cite{Denner:2009gj}.  We also do
not include the (loop-induced) contributions of the partonic process
$\Pg\Pg\to\Pnl\Pnlb\,\Pg$ which can be estimated to be below one percent
based on \citere{vanderBij:1988ac}.

While the virtual QCD corrections consist of up to box (4-point)
diagrams only, the more complicated NLO EW corrections, shown in
\reffi{fi:EW_VF_qqZg}, involve also pentagon (5-point) diagrams (see
\reffi{fi:EW_pent_qqZg}).  As in our earlier work on
$\Plp\Plm+\mathrm{jet}$ production, we follow the traditional
Feynman-dia\-gram\-ma\-tic approach and evaluate tensor and scalar
one-loop integrals (up to pentagon diagrams for the EW corrections) with
complex masses using the methods and results of
\citeres{Passarino:1979jh} and \cite{'tHooft:1979xw}, respectively.

\bfi
\begin{center}
\unitlength=2.5bp%

\begin{small}
\begin{feynartspicture}(160,40)(4,1)
\FALabel(-5,20)[l]{Self-energy insertions:}
\FADiagram{}
\FAProp(0.,15.)(4.,10.)(0.,){/Straight}{1}
\FALabel(1.26965,12.0117)[tr]{$q$}
\FAProp(0.,5.)(4.,10.)(0.,){/Cycles}{0}
\FALabel(2.73035,7.01172)[tl]{$\mathrm{g}$}
\FAProp(20.,17.)(15.5,13.5)(0.,){/Straight}{-1}
\FALabel(17.2784,15.9935)[br]{$\nu_l$}
\FAProp(20.,10.)(15.5,13.5)(0.,){/Straight}{1}
\FALabel(18.2216,12.4935)[bl]{$\nu_l$}
\FAProp(20.,3.)(12.,10.)(0.,){/Straight}{-1}
\FALabel(16.5407,7.18649)[bl]{$\text{q}$}
\FAProp(8.,10.)(4.,10.)(0.,){/Straight}{-1}
\FALabel(6.,11.07)[b]{$q$}
\FAProp(8.,10.)(12.,10.)(0.,){/Straight}{1}
\FALabel(10.,8.93)[t]{$q$}
\FAProp(15.5,13.5)(12.,10.)(0.,){/Sine}{0}
\FALabel(13.134,12.366)[br]{$\mathrm{Z}$}
\FAVert(4.,10.){0}
\FAVert(15.5,13.5){0}
\FAVert(12.,10.){0}
\FAVert(8.,10.){-1}

\FADiagram{}
\FAProp(0.,15.)(3.5,10.)(0.,){/Straight}{1}
\FALabel(0.960191,12.0911)[tr]{$q$}
\FAProp(0.,5.)(3.5,10.)(0.,){/Cycles}{0}
\FALabel(2.53981,7.09113)[tl]{$\mathrm{g}$}
\FAProp(20.,17.)(15.5,13.5)(0.,){/Straight}{-1}
\FALabel(17.2784,15.9935)[br]{$\nu_l$}
\FAProp(20.,10.)(15.5,13.5)(0.,){/Straight}{1}
\FALabel(18.2216,12.4935)[bl]{$\nu_l$}
\FAProp(20.,3.)(9.,10.)(0.,){/Straight}{-1}
\FALabel(14.1478,5.67232)[tr]{$\text{q}$}
\FAProp(12.25,11.75)(15.5,13.5)(0.,){/Sine}{0}
\FALabel(14.1299,11.7403)[tl]{$\mathrm{Z}$}
\FAProp(12.25,11.75)(9.,10.)(0.,){/Sine}{0}
\FALabel(9.8701,11.7597)[br]{$\mathrm{Z}/\gamma$}
\FAProp(3.5,10.)(9.,10.)(0.,){/Straight}{1}
\FALabel(6.25,8.93)[t]{$q$}
\FAVert(3.5,10.){0}
\FAVert(15.5,13.5){0}
\FAVert(9.,10.){0}
\FAVert(12.25,11.75){-1}

\FADiagram{}
\FAProp(0.,15.)(10.,13.5)(0.,){/Straight}{1}
\FALabel(5.22993,15.3029)[b]{$q$}
\FAProp(0.,5.)(10.,4.5)(0.,){/Cycles}{0}
\FALabel(4.9226,3.68193)[t]{$\mathrm{g}$}
\FAProp(20.,17.)(16.,13.5)(0.,){/Straight}{-1}
\FALabel(17.4593,15.9365)[br]{$\nu_l$}
\FAProp(20.,10.)(16.,13.5)(0.,){/Straight}{1}
\FALabel(18.5407,12.4365)[bl]{$\nu_l$}
\FAProp(20.,3.)(10.,4.5)(0.,){/Straight}{-1}
\FALabel(15.2299,4.80285)[b]{$\text{q}$}
\FAProp(10.,9.)(10.,13.5)(0.,){/Straight}{-1}
\FALabel(8.93,11.25)[r]{$\text{q}$}
\FAProp(10.,9.)(10.,4.5)(0.,){/Straight}{1}
\FALabel(8.93,6.75)[r]{$\text{q}$}
\FAProp(10.,13.5)(16.,13.5)(0.,){/Sine}{0}
\FALabel(13.,14.57)[b]{$\mathrm{Z}$}
\FAVert(10.,13.5){0}
\FAVert(10.,4.5){0}
\FAVert(16.,13.5){0}
\FAVert(10.,9.){-1}

\FADiagram{}
\FAProp(0.,15.)(9.,13.5)(0.,){/Straight}{1}
\FALabel(4.75482,15.2989)[b]{$q$}
\FAProp(0.,5.)(9.,5.5)(0.,){/Cycles}{0}
\FALabel(4.58598,4.18239)[t]{$\mathrm{g}$}
\FAProp(20.,17.)(16.,13.5)(0.,){/Straight}{-1}
\FALabel(17.4593,15.9365)[br]{$\nu_l$}
\FAProp(20.,10.)(16.,13.5)(0.,){/Straight}{1}
\FALabel(18.5407,12.4365)[bl]{$\nu_l$}
\FAProp(20.,3.)(9.,5.5)(0.,){/Straight}{-1}
\FALabel(14.8435,5.28146)[b]{$\text{q}$}
\FAProp(12.5,13.5)(9.,13.5)(0.,){/Sine}{0}
\FALabel(9.75,15.27)[b]{$\mathrm{Z}/\gamma$}
\FAProp(12.5,13.5)(16.,13.5)(0.,){/Sine}{0}
\FALabel(14.25,11.43)[t]{$\mathrm{Z}$}
\FAProp(9.,13.5)(9.,5.5)(0.,){/Straight}{1}
\FALabel(7.93,9.5)[r]{$\text{q}$}
\FAVert(9.,13.5){0}
\FAVert(9.,5.5){0}
\FAVert(16.,13.5){0}
\FAVert(12.5,13.5){-1}
\end{feynartspicture}

\begin{feynartspicture}(120,80)(3,2)
\FALabel(-16,42.5)[l]{Triangle insertions:}
\FADiagram{}
\FAProp(0.,15.)(5.5,10.)(0.,){/Straight}{1}
\FALabel(2.18736,11.8331)[tr]{$q$}
\FAProp(0.,5.)(5.5,10.)(0.,){/Cycles}{0}
\FALabel(3.31264,6.83309)[tl]{$\mathrm{g}$}
\FAProp(20.,17.)(15.5,13.5)(0.,){/Straight}{-1}
\FALabel(17.2784,15.9935)[br]{$\nu_l$}
\FAProp(20.,10.)(15.5,13.5)(0.,){/Straight}{1}
\FALabel(18.2216,12.4935)[bl]{$\nu_l$}
\FAProp(20.,3.)(12.,10.)(0.,){/Straight}{-1}
\FALabel(15.4593,5.81351)[tr]{$\text{q}$}
\FAProp(5.5,10.)(12.,10.)(0.,){/Straight}{1}
\FALabel(8.75,8.93)[t]{$q$}
\FAProp(15.5,13.5)(12.,10.)(0.,){/Sine}{0}
\FALabel(13.134,12.366)[br]{$\mathrm{Z}$}
\FAVert(15.5,13.5){0}
\FAVert(12.,10.){0}
\FAVert(5.5,10.){-1}

\FADiagram{}
\FAProp(0.,15.)(5.5,10.)(0.,){/Straight}{1}
\FALabel(2.18736,11.8331)[tr]{$q$}
\FAProp(0.,5.)(5.5,10.)(0.,){/Cycles}{0}
\FALabel(3.31264,6.83309)[tl]{$\mathrm{g}$}
\FAProp(20.,17.)(15.5,13.5)(0.,){/Straight}{-1}
\FALabel(17.2784,15.9935)[br]{$\nu_l$}
\FAProp(20.,10.)(15.5,13.5)(0.,){/Straight}{1}
\FALabel(18.2216,12.4935)[bl]{$\nu_l$}
\FAProp(20.,3.)(12.,10.)(0.,){/Straight}{-1}
\FALabel(15.4593,5.81351)[tr]{$\text{q}$}
\FAProp(5.5,10.)(12.,10.)(0.,){/Straight}{1}
\FALabel(8.75,8.93)[t]{$q$}
\FAProp(15.5,13.5)(12.,10.)(0.,){/Sine}{0}
\FALabel(13.134,12.366)[br]{$\mathrm{Z}$}
\FAVert(5.5,10.){0}
\FAVert(15.5,13.5){0}
\FAVert(12.,10.){-1}

\FADiagram{}
\FAProp(0.,15.)(5.5,10.)(0.,){/Straight}{1}
\FALabel(2.18736,11.8331)[tr]{$q$}
\FAProp(0.,5.)(5.5,10.)(0.,){/Cycles}{0}
\FALabel(3.31264,6.83309)[tl]{$\mathrm{g}$}
\FAProp(20.,17.)(15.5,13.5)(0.,){/Straight}{-1}
\FALabel(17.2784,15.9935)[br]{$\nu_l$}
\FAProp(20.,10.)(15.5,13.5)(0.,){/Straight}{1}
\FALabel(18.2216,12.4935)[bl]{$\nu_l$}
\FAProp(20.,3.)(12.,10.)(0.,){/Straight}{-1}
\FALabel(15.4593,5.81351)[tr]{$\text{q}$}
\FAProp(5.5,10.)(12.,10.)(0.,){/Straight}{1}
\FALabel(8.75,8.93)[t]{$q$}
\FAProp(15.5,13.5)(12.,10.)(0.,){/Sine}{0}
\FALabel(12.434,12.166)[br]{$\mathrm{Z}/\gamma$}
\FAVert(5.5,10.){0}
\FAVert(12.,10.){0}
\FAVert(15.5,13.5){-1}


\FADiagram{}
\FAProp(0.,15.)(10.,13.)(0.,){/Straight}{1}
\FALabel(5.30398,15.0399)[b]{$q$}
\FAProp(0.,5.)(10.,5.5)(0.,){/Cycles}{0}
\FALabel(5.0774,4.18193)[t]{$\mathrm{g}$}
\FAProp(20.,17.)(15.5,13.5)(0.,){/Straight}{-1}
\FALabel(17.2784,15.9935)[br]{$\nu_l$}
\FAProp(20.,10.)(15.5,13.5)(0.,){/Straight}{1}
\FALabel(18.2216,12.4935)[bl]{$\nu_l$}
\FAProp(20.,3.)(10.,5.5)(0.,){/Straight}{-1}
\FALabel(15.3759,5.27372)[b]{$\text{q}$}
\FAProp(10.,13.)(10.,5.5)(0.,){/Straight}{1}
\FALabel(8.93,9.25)[r]{$\text{q}$}
\FAProp(10.,13.)(15.5,13.5)(0.,){/Sine}{0}
\FALabel(12.8903,12.1864)[t]{$\mathrm{Z}$}
\FAVert(10.,13.){0}
\FAVert(15.5,13.5){0}
\FAVert(10.,5.5){-1}



\FADiagram{}
\FAProp(0.,15.)(10.,13.)(0.,){/Straight}{1}
\FALabel(5.30398,15.0399)[b]{$q$}
\FAProp(0.,5.)(10.,5.5)(0.,){/Cycles}{0}
\FALabel(5.0774,4.18193)[t]{$\mathrm{g}$}
\FAProp(20.,17.)(15.5,13.5)(0.,){/Straight}{-1}
\FALabel(17.2784,15.9935)[br]{$\nu_l$}
\FAProp(20.,10.)(15.5,13.5)(0.,){/Straight}{1}
\FALabel(18.2216,12.4935)[bl]{$\nu_l$}
\FAProp(20.,3.)(10.,5.5)(0.,){/Straight}{-1}
\FALabel(15.3759,5.27372)[b]{$\text{q}$}
\FAProp(10.,13.)(10.,5.5)(0.,){/Straight}{1}
\FALabel(8.93,9.25)[r]{$\text{q}$}
\FAProp(10.,13.)(15.5,13.5)(0.,){/Sine}{0}
\FALabel(12.8903,12.1864)[t]{$\mathrm{Z}$}
\FAVert(10.,5.5){0}
\FAVert(15.5,13.5){0}
\FAVert(10.,13.){-1}

\FADiagram{}
\FAProp(0.,15.)(10.,13.)(0.,){/Straight}{1}
\FALabel(5.30398,15.0399)[b]{$q$}
\FAProp(0.,5.)(10.,5.5)(0.,){/Cycles}{0}
\FALabel(5.0774,4.18193)[t]{$\mathrm{g}$}
\FAProp(20.,17.)(15.5,13.5)(0.,){/Straight}{-1}
\FALabel(17.2784,15.9935)[br]{$\nu_l$}
\FAProp(20.,10.)(15.5,13.5)(0.,){/Straight}{1}
\FALabel(18.2216,12.4935)[bl]{$\nu_l$}
\FAProp(20.,3.)(10.,5.5)(0.,){/Straight}{-1}
\FALabel(15.3759,5.27372)[b]{$\text{q}$}
\FAProp(10.,13.)(10.,5.5)(0.,){/Straight}{1}
\FALabel(8.93,9.25)[r]{$\text{q}$}
\FAProp(10.,13.)(15.5,13.5)(0.,){/Sine}{0}
\FALabel(12.5903,12.1864)[t]{$\mathrm{Z}/\gamma$}
\FAVert(10.,13.){0}
\FAVert(10.,5.5){0}
\FAVert(15.5,13.5){-1}

\end{feynartspicture}

\begin{feynartspicture}(160,40)(4,1)
\FALabel(-5,22)[l]{Box and pentagon insertions:}
\FADiagram{}
\FAProp(0.,15.)(5.5,10.)(0.,){/Straight}{1}
\FALabel(2.18736,11.8331)[tr]{$q$}
\FAProp(0.,5.)(5.5,10.)(0.,){/Cycles}{0}
\FALabel(3.31264,6.83309)[tl]{$\mathrm{g}$}
\FAProp(20.,17.)(12,10)(0.,){/Straight}{-1}
\FALabel(17.2784,15.9935)[br]{$\nu_l$}
\FAProp(20.,10.)(12,10)(0.,){/Straight}{1}
\FALabel(18.2216,10.4935)[bl]{$\nu_l$}
\FAProp(20.,3.)(12.,10.)(0.,){/Straight}{-1}
\FALabel(15.4593,5.81351)[tr]{$\text{q}$}
\FAProp(5.5,10.)(12.,10.)(0.,){/Straight}{1}
\FALabel(8.75,8.93)[t]{$q$}
\FAVert(5.5,10.){0}
\FAVert(12.,10.){-1}

\FADiagram{}
\FAProp(0.,15.)(10.,13.)(0.,){/Straight}{1}
\FALabel(5.30398,15.0399)[b]{$q$}
\FAProp(0.,5.)(10.,5.5)(0.,){/Cycles}{0}
\FALabel(5.0774,4.18193)[t]{$\mathrm{g}$}
\FAProp(20.,17.)(10.,13.)(0.,){/Straight}{-1}
\FALabel(17.2784,15.9935)[br]{$\nu_l$}
\FAProp(20.,10.)(10,13.)(0.,){/Straight}{1}
\FALabel(17.2216,11.4935)[bl]{$\nu_l$}
\FAProp(20.,3.)(10.,5.5)(0.,){/Straight}{-1}
\FALabel(15.3759,5.27372)[b]{$\text{q}$}
\FAProp(10.,13.)(10.,5.5)(0.,){/Straight}{1}
\FALabel(8.93,9.25)[r]{$\text{q}$}
\FAVert(10.,13.){0}
\FAVert(10.,5.5){0}
\FAVert(10.,13.){-1}

\FADiagram{}
\FAProp(0.,15.)(10,10.)(0.,){/Straight}{1}
\FALabel(4.18736,11.8331)[tr]{$q$}
\FAProp(0.,5.)(10,10.)(0.,){/Cycles}{0}
\FALabel(3.31264,5.83309)[tl]{$\mathrm{g}$}
\FAProp(20.,17.)(15.5,13.5)(0.,){/Straight}{-1}
\FALabel(17.2784,15.9935)[br]{$\nu_l$}
\FAProp(20.,10.)(15.5,13.5)(0.,){/Straight}{1}
\FALabel(18.2216,12.4935)[bl]{$\nu_l$}
\FAProp(20.,3.)(10,10.)(0.,){/Straight}{-1}
\FALabel(13,5)[tl]{$\text{q}$}
\FAProp(15.5,13.5)(10,10.)(0.,){/Sine}{0}
\FALabel(13.134,12.366)[br]{$\mathrm{Z}$}
\FAVert(15.5,13.5){0}
\FAVert(10,10.){-1}

\FADiagram{}
\FAProp(0.,15.)(10,10.)(0.,){/Straight}{1}
\FALabel(4.18736,11.8331)[tr]{$q$}
\FAProp(0.,5.)(10,10.)(0.,){/Cycles}{0}
\FALabel(3.31264,5.83309)[tl]{$\mathrm{g}$}
\FAProp(20,3)(10,10)(0,){/Straight}{-1}
\FALabel(13,5)[tl]{$\text{q}$}
\FAProp(20.,17.)(10,10)(0.,){/Straight}{-1}
\FALabel(17.4593,15.9365)[br]{$\nu_l$}
\FAProp(20.,10.)(10,10)(0.,){/Straight}{1}
\FALabel(17.5407,10.4365)[bl]{$\nu_l$}
\FAVert(10.,10.){-1}

\end{feynartspicture}

%

\end{small}

\vspace*{-2em}
\end{center}
\mycaption{\label{fi:EW_VF_qqZg} Contributions of different one-particle
  irreducible vertex functions (indicated as blobs) to the LO process
  \refeq{eq:proc2}; there are contributions from self-energies,
  triangles, boxes, and pentagon graphs.} 
\efi 
\bfi
\begin{center}
\unitlength=3bp%

\begin{center}
\begin{small}
\begin{feynartspicture}(120,30)(3,1)

\FADiagram{}
\FAProp(0.,15.)(5.,13.)(0.,){/Straight}{1}
\FALabel(3.07566,14.9591)[b]{$q$}
\FAProp(0.,5.)(5.,7.)(0.,){/Cycles}{0}
\FALabel(3.07566,5.04086)[t]{$g$}
\FAProp(20.,17.)(10.5,14.5)(0.,){/Straight}{-1}
\FALabel(14.8555,16.769)[b]{$\nu_l$}
\FAProp(20.,10.)(14.,10.)(0.,){/Straight}{1}
\FALabel(17.,11.07)[b]{$\nu_l$}
\FAProp(20.,3.)(10.5,5.5)(0.,){/Straight}{-1}
\FALabel(14.8555,3.23103)[t]{$q$}
\FAProp(5.,13.)(5.,7.)(0.,){/Straight}{1}
\FALabel(3.93,10.)[r]{$q$}
\FAProp(5.,13.)(10.5,14.5)(0.,){/Sine}{0}
\FALabel(7.34217,14.7654)[b]{$\PZ$}
\FAProp(5.,7.)(10.5,5.5)(0.,){/Straight}{1}
\FALabel(7.34217,5.23462)[t]{$q$}
\FAProp(10.5,14.5)(14.,10.)(0.,){/Straight}{-1}
\FALabel(12.9935,12.7216)[bl]{$\nu_l$}
\FAProp(14.,10.)(10.5,5.5)(0.,){/Sine}{0}
\FALabel(12.9935,7.27839)[tl]{$\PZ$}
\FAVert(5.,13.){0}
\FAVert(5.,7.){0}
\FAVert(10.5,14.5){0}
\FAVert(14.,10.){0}
\FAVert(10.5,5.5){0}

\FADiagram{}
\FAProp(0.,15.)(5.,13.)(0.,){/Straight}{1}
\FALabel(3.07566,14.9591)[b]{$q$}
\FAProp(0.,5.)(5.,7.)(0.,){/Cycles}{0}
\FALabel(3.07566,5.04086)[t]{$g$}
\FAProp(20.,17.)(10.5,14.5)(0.,){/Straight}{-1}
\FALabel(14.8555,16.769)[b]{$\nu_l$}
\FAProp(20.,10.)(14.,10.)(0.,){/Straight}{1}
\FALabel(17.,11.07)[b]{$\nu_l$}
\FAProp(20.,3.)(10.5,5.5)(0.,){/Straight}{-1}
\FALabel(14.8555,3.23103)[t]{$q$}
\FAProp(5.,13.)(5.,7.)(0.,){/Straight}{1}
\FALabel(3.93,10.)[r]{$q'$}
\FAProp(5.,13.)(10.5,14.5)(0.,){/Sine}{-1}
\FALabel(7.34217,14.7654)[b]{$\PW$}
\FAProp(5.,7.)(10.5,5.5)(0.,){/Straight}{1}
\FALabel(7.34217,5.23462)[t]{$q'$}
\FAProp(10.5,14.5)(14.,10.)(0.,){/Straight}{-1}
\FALabel(12.9935,12.7216)[bl]{$l$}
\FAProp(14.,10.)(10.5,5.5)(0.,){/Sine}{-1}
\FALabel(12.9935,7.27839)[tl]{$\PW$}
\FAVert(5.,13.){0}
\FAVert(5.,7.){0}
\FAVert(10.5,14.5){0}
\FAVert(14.,10.){0}
\FAVert(10.5,5.5){0}

\FADiagram{}
\FAProp(0.,15.)(5.,13.)(0.,){/Straight}{1}
\FALabel(3.07566,14.9591)[b]{$q$}
\FAProp(0.,5.)(5.,7.)(0.,){/Cycles}{0}
\FALabel(3.07566,5.04086)[t]{$g$}
\FAProp(20.,17.)(14.,10.)(0.,){/Straight}{-1}
\FALabel(18.2902,16.1827)[br]{$\nu_l$}
\FAProp(20.,10.)(10.5,14.5)(0.,){/Straight}{1}
\FALabel(18.373,9.75407)[tr]{$\nu_l$}
\FAProp(20.,3.)(10.5,5.5)(0.,){/Straight}{-1}
\FALabel(14.8555,3.23103)[t]{$q$}
\FAProp(5.,13.)(5.,7.)(0.,){/Straight}{1}
\FALabel(3.93,10.)[r]{$q$}
\FAProp(5.,13.)(10.5,14.5)(0.,){/Sine}{0}
\FALabel(7.34217,14.7654)[b]{$\PZ$}
\FAProp(5.,7.)(10.5,5.5)(0.,){/Straight}{1}
\FALabel(7.34217,5.23462)[t]{$q$}
\FAProp(14.,10.)(10.5,14.5)(0.,){/Straight}{-1}
\FALabel(11.5065,11.7784)[tr]{$\nu_l$}
\FAProp(14.,10.)(10.5,5.5)(0.,){/Sine}{0}
\FALabel(12.9935,7.27839)[tl]{$\PZ$}
\FAVert(5.,13.){0}
\FAVert(5.,7.){0}
\FAVert(14.,10.){0}
\FAVert(10.5,14.5){0}
\FAVert(10.5,5.5){0}

\end{feynartspicture}
\end{small}
\end{center}

\vspace*{-2em}
\end{center}
\mycaption{\label{fi:EW_pent_qqZg} Virtual pentagon
contributions to the process \refeq{eq:proc2}. Note that for external
bottom quarks the exchange of two \PW\ bosons leads to diagrams with
massive top-quark lines ($q'=\mathrm{top}$) in the loop.}  
\efi

The partonic processes with (anti-)bottom quarks in the initial state
involve massive top quarks in EW loop diagrams.  These affect the total
EW correction by about a permille for the most inclusive cross section
discussed in \refse{se:numres} at the 14~TeV LHC and by even less at
lower hadronic CM energy or at the Tevatron. For less inclusive cross
sections when the EW Sudakov logarithms dominate at high partonic CM
energy, the top mass does not play a role.

\subsection{Real corrections}
\label{se:real} 

Compared to the $\Plp\Plm$ final state discussed
in~\citere{Denner:2011vu}, the EW corrections are substantially
simpler, since final-state photon radiation off leptons is absent in
the monojet production scenario.  The emission of an additional photon
in the partonic processes \refeq{eq:proc1}--\refeq{eq:proc3} leads to
the processes
\begin{eqnarray}
\label{eq:bremsproc1}%
& \Pq_i \, \, \Pqbar_i & \to \Pnl \Pnlb \, \Pg \, \gamma \, ,\\
\label{eq:bremsproc2}%
& \Pq_i \, \, \Pg      & \to \Pnl \Pnlb \, \Pq_i \, \gamma \, ,\\
\label{eq:bremsproc3}%
& \Pqbar_i \, \, \Pg   & \to \Pnl \Pnlb \, \Pqbar_i \, \gamma \, .
\end{eqnarray}
For the process \refeq{eq:bremsproc2} the relevant Feynman diagrams are
listed in \reffi{fi:EW_real_qgZgammag}.  The corresponding amplitudes
may be obtained from the amplitudes for the
$\Plp\Plm+\mathrm{jet}+\gamma$ final state by switching off the virtual
photon as well as the final-state radiation off the leptons.  In
addition one has to replace the $\PZ\Plm\Plp$ coupling by the
$\PZ\Pnl\Pnlb$ coupling.

In order to treat the soft and collinear singularities in an efficient
way we use the dipole subtraction formalism as specified for photon
emission in \citeres{Dittmaier:1999mb,Dittmaier:2008md}.

\bfi
\begin{center}
\unitlength=3bp%

\begin{center}
\begin{small}
\begin{feynartspicture}(150,60)(3,2)

\FADiagram{}
\FAProp(0.,15.)(4.,10.)(0.,){/Straight}{1}
\FALabel(1.26965,12.0117)[tr]{$q$}
\FAProp(0.,5.)(4.,10.)(0.,){/Cycles}{0}
\FALabel(2.73035,7.01172)[tl]{$\Pg$}
\FAProp(20.,18.)(14.,15.)(0.,){/Straight}{-1}
\FALabel(16.7868,17.4064)[br]{$\nu_l$}
\FAProp(20.,12.)(14.,15.)(0.,){/Straight}{1}
\FALabel(16.7868,12.5936)[tr]{$\nu_l$}
\FAProp(20.,8.)(14.,5.)(0.,){/Straight}{-1}
\FALabel(16.7868,7.40636)[br]{$q$}
\FAProp(20.,2.)(14.,5.)(0.,){/Sine}{0}
\FALabel(16.7868,2.59364)[tr]{$\gamma$}
\FAProp(4.,10.)(10.,10.)(0.,){/Straight}{1}
\FALabel(7.,11.07)[b]{$q$}
\FAProp(14.,15.)(10.,10.)(0.,){/Sine}{0}
\FALabel(11.2697,12.9883)[br]{$\PZ$}
\FAProp(14.,5.)(10.,10.)(0.,){/Straight}{-1}
\FALabel(11.2697,7.01172)[tr]{$q$}
\FAVert(4.,10.){0}
\FAVert(14.,15.){0}
\FAVert(14.,5.){0}
\FAVert(10.,10.){0}

\FADiagram{}
\FAProp(0.,15.)(4.,10.)(0.,){/Straight}{1}
\FALabel(2.73035,12.9883)[bl]{$q$}
\FAProp(0.,5.)(4.,10.)(0.,){/Cycles}{0}
\FALabel(0.723045,8.42556)[br]{$\Pg$}
\FAProp(20.,18.)(15.5,15.)(0.,){/Straight}{-1}
\FALabel(17.3702,17.3097)[br]{$\nu_l$}
\FAProp(20.,12.)(15.5,15.)(0.,){/Straight}{1}
\FALabel(17.3702,12.6903)[tr]{$\nu_l$}
\FAProp(20.,8.)(11.5,12.5)(0.,){/Straight}{-1}
\FALabel(15.5048,9.36013)[tr]{$q$}
\FAProp(20.,2.)(8.,9.)(0.,){/Sine}{0}
\FALabel(13.699,4.64114)[tr]{$\gamma$}
\FAProp(4.,10.)(8.,9.)(0.,){/Straight}{1}
\FALabel(5.62407,8.47628)[t]{$q$}
\FAProp(15.5,15.)(11.5,12.5)(0.,){/Sine}{0}
\FALabel(13.1585,14.5844)[br]{$\PZ$}
\FAProp(11.5,12.5)(8.,9.)(0.,){/Straight}{-1}
\FALabel(9.13398,11.366)[br]{$q$}
\FAVert(4.,10.){0}
\FAVert(15.5,15.){0}
\FAVert(11.5,12.5){0}
\FAVert(8.,9.){0}

\FADiagram{}
\FAProp(0.,15.)(10.,5.5)(0.,){/Straight}{1}
\FALabel(3.19219,13.2012)[bl]{$q$}
\FAProp(0.,5.)(10.,14.5)(0.,){/Cycles}{0}
\FALabel(3.19219,6.79875)[tl]{$\Pg$}
\FAProp(20.,18.)(17.,15.)(0.,){/Straight}{-1}
\FALabel(17.884,17.116)[br]{$\nu_l$}
\FAProp(20.,12.)(17.,15.)(0.,){/Straight}{1}
\FALabel(19.116,14.116)[bl]{$\nu_l$}
\FAProp(20.,8.)(13.5,14.)(0.,){/Straight}{-1}
\FALabel(16.1787,10.3411)[tr]{$q$}
\FAProp(20.,2.)(10.,5.5)(0.,){/Sine}{0}
\FALabel(14.488,2.76702)[t]{$\gamma$}
\FAProp(10.,5.5)(10.,14.5)(0.,){/Straight}{1}
\FALabel(11.27,8.5)[l]{$q$}
\FAProp(10.,14.5)(13.5,14.)(0.,){/Straight}{1}
\FALabel(11.9692,15.3044)[b]{$q$}
\FAProp(17.,15.)(13.5,14.)(0.,){/Sine}{0}
\FALabel(14.8242,15.5104)[b]{$\PZ$}
\FAVert(10.,5.5){0}
\FAVert(10.,14.5){0}
\FAVert(17.,15.){0}
\FAVert(13.5,14.){0}

\FADiagram{}
\FAProp(0.,15.)(8.,14.5)(0.,){/Straight}{1}
\FALabel(4.09669,15.817)[b]{$q$}
\FAProp(0.,5.)(8.,5.5)(0.,){/Cycles}{0}
\FALabel(4.09669,4.18302)[t]{$\Pg$}
\FAProp(20.,18.)(14.,15.)(0.,){/Straight}{-1}
\FALabel(16.7868,17.4064)[br]{$\nu_l$}
\FAProp(20.,12.)(14.,15.)(0.,){/Straight}{1}
\FALabel(16.7868,12.5936)[tr]{$\nu_l$}
\FAProp(20.,8.)(14.,5.)(0.,){/Straight}{-1}
\FALabel(16.7868,7.40636)[br]{$q$}
\FAProp(20.,2.)(14.,5.)(0.,){/Sine}{0}
\FALabel(16.7868,2.59364)[tr]{$\gamma$}
\FAProp(8.,14.5)(8.,5.5)(0.,){/Straight}{1}
\FALabel(6.93,10.)[r]{$q$}
\FAProp(8.,14.5)(14.,15.)(0.,){/Sine}{0}
\FALabel(10.8713,15.8146)[b]{$\PZ$}
\FAProp(8.,5.5)(14.,5.)(0.,){/Straight}{1}
\FALabel(10.8713,4.18535)[t]{$q$}
\FAVert(8.,14.5){0}
\FAVert(8.,5.5){0}
\FAVert(14.,15.){0}
\FAVert(14.,5.){0}

\FADiagram{}
\FAProp(0.,15.)(8.,14.5)(0.,){/Straight}{1}
\FALabel(4.09669,15.817)[b]{$q$}
\FAProp(0.,5.)(8.,5.5)(0.,){/Cycles}{0}
\FALabel(4.09669,4.18302)[t]{$\Pg$}
\FAProp(20.,18.)(13.5,15.)(0.,){/Straight}{-1}
\FALabel(16.5805,17.4273)[br]{$\nu_l$}
\FAProp(20.,12.)(13.5,15.)(0.,){/Straight}{1}
\FALabel(16.5805,12.5727)[tr]{$\nu_l$}
\FAProp(20.,8.)(8.,5.5)(0.,){/Straight}{-1}
\FALabel(15.437,8.0267)[b]{$q$}
\FAProp(20.,2.)(8.,10.)(0.,){/Sine}{0}
\FALabel(16.0956,3.41321)[tr]{$\gamma$}
\FAProp(8.,14.5)(13.5,15.)(0.,){/Sine}{0}
\FALabel(10.6097,15.8136)[b]{$\PZ$}
\FAProp(8.,14.5)(8.,10.)(0.,){/Straight}{1}
\FALabel(6.93,12.25)[r]{$q$}
\FAProp(8.,5.5)(8.,10.)(0.,){/Straight}{-1}
\FALabel(6.93,7.75)[r]{$q$}
\FAVert(8.,14.5){0}
\FAVert(8.,5.5){0}
\FAVert(13.5,15.){0}
\FAVert(8.,10.){0}

\FADiagram{}
\FAProp(0.,15.)(8.,14.5)(0.,){/Straight}{1}
\FALabel(4.09669,15.817)[b]{$q$}
\FAProp(0.,5.)(8.,5.5)(0.,){/Cycles}{0}
\FALabel(4.09669,4.18302)[t]{$\Pg$}
\FAProp(20.,18.)(15.5,15.)(0.,){/Straight}{-1}
\FALabel(17.3702,17.3097)[br]{$\nu_l$}
\FAProp(20.,12.)(15.5,15.)(0.,){/Straight}{1}
\FALabel(17.3702,12.6903)[tr]{$\nu_l$}
\FAProp(20.,8.)(8.,5.5)(0.,){/Straight}{-1}
\FALabel(11.551,4.73525)[t]{$q$}
\FAProp(20.,2.)(8.,14.5)(0.,){/Sine}{0}
\FALabel(16.7997,3.80687)[tr]{$\gamma$}
\FAProp(8.,14.5)(8.,10.)(0.,){/Straight}{1}
\FALabel(6.93,12.25)[r]{$q$}
\FAProp(8.,5.5)(8.,10.)(0.,){/Straight}{-1}
\FALabel(6.93,7.75)[r]{$q$}
\FAProp(15.5,15.)(8.,10.)(0.,){/Sine}{0}
\FALabel(13.2061,14.5881)[br]{$\PZ$}
\FAVert(8.,14.5){0}
\FAVert(8.,5.5){0}
\FAVert(15.5,15.){0}
\FAVert(8.,10.){0}

\end{feynartspicture}
\end{small}
\end{center}

\vspace*{-2em}
\end{center}
\mycaption{\label{fi:EW_real_qgZgammag}
Real photonic bremsstrahlung corrections to the LO process \refeq{eq:proc2}.}
\efi

As in all processes with jets in the final state, the inclusion of EW
corrections to monojet production asks for a precise event definition in
order to distinguish single-jet from single-photon production. We follow
the strategy used for $\Pnl\Plp + \mathrm{jet}$ and $\Plm\Plp +
\mathrm{jet}$ production which is detailed in
\citeres{Denner:2011vu,Denner:2010ia,Denner:2009gj}: We exclude jets
which primarily consist of a hard photon (see \refse{se:numres}) and
capture the non-perturbative physics in the collinear quark--photon
splittings~\cite{Glover:1993xc} by means of the measured fragmentation
function~\cite{Buskulic:1995au}.  The error on our predictions
associated to the uncertainties in the fragmentation function are
negligible.

The calculation of the QCD corrections to single-jet production is
completely analogous to $\mathrm{dilepton}+\mathrm{jet}$ production and
has been discussed in Section 2.3 of~\citere{Denner:2011vu}. 
It contains a plethora of partonic channels (see Eqs.~(2.11)--(2.21) in
Ref.~\cite{Denner:2011vu}), in particular involving six-fermion
processes with a rich flavour structure of the involved
\mbox{(anti-)}quarks. For six-fermion processes with identical quarks,
again, we do not consider interference contributions between purely
electroweak diagrams and gluon-exchange diagrams which enter at order
$\mathcal{O}(\alpha^3\alphas)$. These contributions are
colour-suppressed and do not exhibit any further enhancement. Thus, they
only lead to tiny corrections, as has been demonstrated in
Ref.~\cite{Denner:2009gj}.

\section{Numerical results}
\label{se:numres}

\subsection{Input parameters and setup}
\label{se:SMinput}

The relevant SM input parameters are
\begin{equation}\arraycolsep 2pt
\begin{array}[b]{lcllcllcl}
\GF & = & 1.16637 \times 10^{-5} \GeV^{-2}, \quad&
\alpha_{\mathrm{s}}(\MZ) &=& 0.1202 , 
&&&
\\
\MW^{\OS} & = & 80.398\GeV, &
\Gamma_\PW^{\OS} & = & 2.141\GeV, \\
\MZ^{\OS} & = & 91.1876\GeV, &
\Gamma_\PZ^{\OS} & = & 2.4952\GeV, & & & \\
M_\PH & = & 120\GeV, & m_\Pt & = & 172.6\;\GeV\,,
\end{array}
\label{eq:SMpar}
\end{equation}
which essentially follow \citere{Amsler:2008zzb}. To facilitate
comparisons, we stick to the input parameters used in
\citeres{Denner:2011vu,Denner:2009gj} for the analysis of
$\nu_ll^++\mathrm{jet}$ and $l^+l^-+\mathrm{jet}$ production,
respectively. The numerical results presented in
this paper will only change marginally using up-to-date values of
$\MW$, $\MH$, and $\Mt$. 
The CKM matrix only appears in loops and is set to unity,
because its effect is negligible.

As in our earlier work on $V+\mathrm{jet}$ production,
using the complex-mass scheme~\cite{Denner:2005fg}, we convert 
the ``on-shell'' (OS) values of $M_V^{\OS}$ and $\Ga_V^{\OS}$ 
($V=\PW,\PZ$) into the ``pole values'' \cite{Bardin:1988xt} 
and use the corresponding numerical values
\beqar
\begin{array}[b]{r@{\,}l@{\qquad}r@{\,}l}
\MW &= 80.370\ldots\GeV, & \GW &= 2.1402\ldots\GeV, \\
\MZ &= 91.153\ldots\GeV,& \GZ &= 2.4943\ldots\GeV
\label{eq:m_ga_pole_num}
\end{array}
\eeqar
in the numerics. However, using $M_V^{\OS}$ instead would be hardly 
visible in the results. We again adopt the
$\GF$ scheme so that the results are practically 
independent of the light fermion masses.

We use the central MSTW2008NLO PDF set~\cite{Martin:2009iq} 
and the $\alpha_{\mathrm{s}}$ running from the LHAPDF 
collaboration~\cite{Whalley:2005nh} 
(implying $\alpha_{\mathrm{s}}(\MZ)$ given in \refeq{eq:SMpar}). 
Only the photon-induced processes are evaluated with the MRSTQED2004 set of
PDFs~\cite{Martin:2004dh} (implying $\alpha_{\mathrm{s}}(\MZ) = 0.1190$).

Numerical results are presented for the identified QCD and QED 
factorization scales and the QCD renormalization scale 
$\mu=\MZ$ or the phase-space-dependent scale 
\begin{equation}
\label{eq:scale_choice}
\mu^{\mathrm{var}}=\sqrt{\MZ^2+(p_{\rT}^\mathrm{had})^2} \,  ,
\end{equation}
where $p_{\rT}^\mathrm{had}$ is
given by the $p_{\rT}$ of the summed four-momenta of all partons
(quarks and/or gluons) in the final state,
which is more adequate for high-$p_{\rT}$ jets 
(see discussion in \citere{Denner:2011vu} or \citere{Bauer:2009km}).

\subsection{Phase-space cuts and event selection}
\label{se:cuts}

We define the monojet signature by the recombination procedure for 
quarks, gluons, and photons into jets and the basic
cuts discussed in the following subsections.

\subsubsection{Recombination}

To define the recombination procedure and the separation cuts, we use
the variables $R_{ij} = \sqrt{(y_{i}-y_{j})^2+\phi_{ij}^2}$, where
$y_{i}$ denotes the rapidity 
$y_i = \frac{1}{2} \ln [(E + p_\RL)/(E - p_\RL)]$ of particle $i$
and $\phi_{ij}$ is the azimuthal angle in the transverse plane between
the particles $i$ and $j$. In the definition of the rapidity, $E$
denotes the particle's energy and $p_{\RL}$ its three-momentum along
the beam axis. The recombination procedure, where we simply add
four-momenta to form a pseudo-particle, works as follows:
\begin{enumerate}
\item A photon and a parton $a$ (quark or gluon) are recombined for
  $R_{\gamma a} < 0.5$. 
  The energy fraction of
  the photon inside the jet $z_{\gamma}= E_{\gamma}/(E_{\gamma}+E_a)$ is used   
  to distinguish between single-jet and single-photon production
  (using quark-to-photon fragmentation functions to describe the
  non-perturbative collinear splitting). For
  $z_{\gamma} > 0.7$ the event is rejected.  
  Our results are not very sensitive to
  the specific choice of the cut on $z_{\gamma}$.
\item Two partons $a,b$ are recombined for $R_{ab} < 0.5$. For our
  simple final-state configurations, this procedure is equivalent to the
  Tevatron Run II $k_{\rT}$ algorithm \cite{Blazey:2000qt}, the
  anti-$k_{\rT}$ algorithm~\cite{Cacciari:2008gp}, and similar
  algorithms for jet reconstruction with resolution parameter
  $D=0.5$.
\end{enumerate}

\subsubsection{Basic cuts}
\label{se:basic_cuts}

After applying the recombination procedure of the previous section we define
monojet events by the following basic cuts:
\begin{enumerate}
\item A partonic object (after a possible recombination) is called a
  jet if its transverse momentum $p_{\rT}$ is larger than
  $p^{\mathrm{cut}}_{\rT,\mathrm{jet}} = 25\GeV$.  Events are required
  to include at least one jet.
\item We demand a missing transverse momentum (i.e.\ the transverse
  momentum of the neutrino pair) $\ptmiss > 25\GeV$. Of
  course, at LO this requirement is automatically fulfilled, since the
  virtual \PZ\ and the jet are always back-to-back.  
\item The events have to
  be central, i.e.\ the jet has to be produced in the
  rapidity range $|y| < y_{\mathrm{max}} = 2.5$.
\item To suppress background from a mis-measurement of the transverse
  momentum of one of the jets in events with two hard back-to-back jets,
  which also lead to a signature with missing momentum, we require a
  separation of a jet and the missing momentum $\ptmiss$ in the
  transverse plane.  As proposed in Ref.~\cite{Aad:2011xw},
  we discard events with
  $\phi_{\mathrm{j}\PZ}<0.5$, where $\phi_{\mathrm{j}\PZ}$ is the angle
  between $p_{\rT,\mathrm{jet}}$ and $\ptmiss$. A similar
  separation cut could be applied for the photon and $\ptmiss$. However,
  we have checked that such a separation only insignificantly changes
  the numerical results.
\end{enumerate} 

Additionally, we present results obtained by applying an explicit veto
against a second hard jet with $p_{\rT} > p_{\rT,\Pj_1}/2$, where
$p_{\rT,\Pj_1}$ denotes the $p_{\rT}$ of the ``leading'' jet, i.e.\ the
one with maximal $p_{\rT}$. This two-jet veto reduces the relative QCD
corrections and moreover enforces the kinematic structure of monojet
events, i.e.\ the transverse momentum of the hard jet is to be (mainly)
balanced by missing transverse momentum and not by a second hard jet
(see discussion below).

\subsection{Results on cross sections and distributions}
\label{se:CSresults}

Numerical results are presented for the production of a neutrino pair 
in association with a jet at the Tevatron 
($\Pp\bar\Pp$~collisions with a 
CM energy of $\sqrt{s}=1.96\TeV$) and at the LHC. 
For the latter, we show results for \Pp\Pp~collisions at
$\sqrt{s}=8\TeV$, corresponding to the available energy in the year 
2012, as well as $\sqrt{s}=14\TeV$, the design energy 
of the LHC.

\subsubsection{Overview of cross-section predictions}

We present the LO cross section $\sigma_0$ and corrections $\de$,
defined relative to the LO cross section by $\sigma =
\sigma_0\times\left(1+\de\right)$.  The EW corrections are labelled
$\delta_\mathrm{EW}^{\mu=\MZ}$ or $\delta_\mathrm{EW}^\mathrm{var}$
indicating a fixed scale choice or the phase-space dependent scale
choice as specified in \refeq{eq:scale_choice}, respectively. For the EW
corrections the difference is small, as expected, while for the QCD part
a sensible scale choice is crucial as discussed below.

As already observed in $\PW+\mathrm{jet}$ and $\Plm\Plp+\mathrm{jet}$
production in Refs.~\cite{Denner:2011vu,Denner:2009gj} and shown for
$\Pnl\Pnlb+\mathrm{jet}$ below, the QCD corrections become larger and
larger with increasing $p_{\rT}$ of the leading jet due to dijet
kinematics where one of the quark lines radiates a relatively soft \PZ\
boson.  The cut on $\phi_{\mathrm{j}\PZ}$, discussed in
Section~\ref{se:basic_cuts}, partly removes these
configurations. However, a usual cut value like
$\phi_{\mathrm{j}\PZ}<0.5$ is too small to avoid the problem. This part
of the cross section does not really correspond to a true NLO correction
to single-jet production. It even contradicts the intuitive
understanding how a monojet event looks like. Fortunately, it can be
easily separated employing the veto introduced at the end of
\refse{se:basic_cuts}.  This particular jet veto yields a sensible
definition of monojet events since it is equivalent to raising the
required amount of missing transverse momentum along with the transverse
momentum of the observed leading jet.  NLO QCD corrections with a jet
veto ($\delta^{\mu=\MZ}_\mathrm{QCD,veto}$,
$\delta^\mathrm{var}_\mathrm{QCD,veto}$) and without a jet veto
($\delta^{\mu=\MZ}_\mathrm{QCD}$, $\delta^\mathrm{var}_\mathrm{QCD}$)
are presented below to demonstrate the importance of the jet veto.

The impact $\delta_{\gamma}$ of the photon-induced tree-level processes
\refeq{eq:proc4} and \refeq{eq:proc5} is shown relative to the LO cross
section at $\mathcal{O}(\alpha^2\alpha_{\mathrm{s}})$ without
initial-state photons.

\begin{table}
                                                                                                                                         $$ \begin{array}{c|rrrrrr}
                                                                  \multicolumn{7}{c}{\Pp\Pp \to \Pnl \Pnlb\; \mathrm{jet} + X \;\mbox{at} \;\sqrt{s} =14 \TeV} \\
              \hline p_{\rT,\mathrm{jet}} / \GeV & 25-\infty \;\;\; & 50-\infty \;\;\; & 100-\infty \;\; & 200-\infty \;\; & 500-\infty \;\; & 1000-\infty \; \\ 
                                                                                                                                                     \hline\hline
\si_0^{\mu = \MZ}/\pba                                       \; & \; 1627.25(3)      \; & \; 582.66(3)       \; & \; 125.669(5)      \; & \; 15.0488(5)      \; & \; 0.40302(1)      \; & \; 0.0121944(2)    \\ 
\si_0^{\mathrm{var}}/\pba                                    \; & \; 1608.99(2)      \; & \; 566.52(2)       \; & \; 116.092(4)      \; & \; 12.2928(4)      \; & \; 0.243365(6)     \; & \; 0.00523456(9)   \\ 
   \hline \hline                                                                                                         
\de_{\EW}^{\mu = \MZ} / \%                                   \; & \; 0.0\phz         \; & \; -0.4\phz        \; & \; -2.1\phz        \; & \; -6.5\phz        \; & \; -16.7\phz       \; & \; -27.6(1)        \\ 
\de_{\EW}^{\mathrm{var}}/\%                                  \; & \; 0.0\phz         \; & \; -0.4\phz        \; & \; -2.0\phz        \; & \; -6.2\phz        \; & \; -16.1\phz       \; & \; -26.6(1)        \\ 
   \hline \hline                                                                                                         
\de_{\QCD}^{\mu = \MZ}/\%                                    \; & \; 8.5\phz         \; & \; 38.5(1)         \; & \; 51.8(1)         \; & \; 76.2(1)         \; & \; 126.4(1)        \; & \; 179.3(1)        \\ 
\de_{\QCD}^{\mathrm{var}}/\%                                 \; & \; 8.4\phz         \; & \; 39.7(1)         \; & \; 57.8(1)         \; & \; 97.2(1)         \; & \; 210.5(1)        \; & \; 397.3(1)        \\ 
  \hline                                                                                                                 
\de_{\QCD,\veto}^{\mu = \MZ}/\%                              \; & \; -8.8\phz        \; & \; 6.9(1)          \; & \; 7.1(1)          \; & \; -0.2(1)         \; & \; -31.4(1)        \; & \; -68.9(1)        \\ 
\de_{\QCD,\veto}^{\mathrm{var}}/\%                           \; & \; -8.0\phz        \; & \; 9.6(1)          \; & \; 14.9(1)         \; & \; 20.2(1)         \; & \; 21.0(1)         \; & \; 23.4(1)         \\ 
   \hline \hline                                                                                                         
\de_{\ga}^{\mu = \MZ}/\%                                     \; & \; 0.0\phz         \; & \; 0.1\phz         \; & \; 0.1\phz         \; & \; 0.1\phz         \; & \; 0.1\phz         \; & \; 0.2\phz         \\ 
\de_{\ga}^{\mathrm{var}}/\%                                  \; & \; 0.0\phz         \; & \; 0.1\phz         \; & \; 0.1\phz         \; & \; 0.1\phz         \; & \; 0.2\phz         \; & \; 0.5\phz         \\ 
   \hline \hline                                                                                                         
\si_{\mathrm{full, veto}}^{\mathrm{var}}/\pba                \; & \; 1480.7(2)       \; & \; 619.2(3)        \; & \; 131.23(5)       \; & \; 14.029(6)       \; & \; 0.2557(1)       \; & \; 0.005090(3)     \\ 
 \end{array} $$

  \mycaption{\label{ta:ptj_LHC} Integrated cross sections for different
    cuts on the $p_{\rT}$ of the leading jet (jet with highest
    $p_{\rT}$) at the LHC with  
    $\sqrt{s}=14\TeV$. We show the LO results both for a variable and 
    for a constant scale along with the corresponding EW corrections 
    $\de_{\EW}$, the QCD corrections $\de_{\QCD}$ with or without employing
    a veto on a second hard jet, and the corrections due to photon-induced  
    processes $\de_{\ga}$. Finally, we show the full NLO cross section
    $\si_{\mathrm{full},\mathrm{veto}}^{\mathrm{var}}$ for which all the
    corrections are added to the LO results for a variable scale.
    The error from the Monte Carlo integration for the last digit(s)
    is given in parenthesis as far as significant. 
  }
\end{table}

\begin{table}[p]
                                                                                                                                       $$ \begin{array}{c|rrrrrr}
                                                                   \multicolumn{7}{c}{\Pp\Pp \to \Pnl \Pnlb\; \mathrm{jet} + X \;\mbox{at} \;\sqrt{s} =8 \TeV} \\
              \hline p_{\rT,\mathrm{jet}} / \GeV & 25-\infty \;\;\; & 50-\infty \;\;\; & 100-\infty \;\; & 200-\infty \;\; & 500-\infty \;\; & 1000-\infty \; \\ 
                                                                                                                                                     \hline\hline
\si_0^{\mu = \MZ}/\pba                                       \; & \; 782.38(1)       \; & \; 263.82(1)       \; & \; 50.954(2)       \; & \; 4.9461(2)       \; & \; 0.075842(2)     \; & \; 0.00086825(2)   \\ 
\si_0^{\mathrm{var}}/\pba                                    \; & \; 768.26(1)       \; & \; 252.79(1)       \; & \; 45.631(2)       \; & \; 3.8100(1)       \; & \; 0.0409508(8)    \; & \; 0.0003054(3)    \\ 
   \hline \hline                                                                                                         
\de_{\EW}^{\mu = \MZ} / \%                                   \; & \; 0.1\phz         \; & \; -0.3\phz        \; & \; -2.0\phz        \; & \; -6.2\phz        \; & \; -16.1\phz       \; & \; -26.9(1)        \\ 
\de_{\EW}^{\mathrm{var}}/\%                                  \; & \; 0.1\phz         \; & \; -0.3\phz        \; & \; -1.8\phz        \; & \; -5.9\phz        \; & \; -15.5\phz       \; & \; -25.8(1)        \\ 
   \hline \hline                                                                                                         
\de_{\QCD}^{\mu = \MZ}/\%                                    \; & \; 8.2\phz         \; & \; 33.6(1)         \; & \; 41.5(1)         \; & \; 58.0(1)         \; & \; 97.2(1)         \; & \; 151.7(1)        \\ 
\de_{\QCD}^{\mathrm{var}}/\%                                 \; & \; 8.8\phz         \; & \; 36.6(1)         \; & \; 51.5(1)         \; & \; 87.6(1)         \; & \; 203.7(1)        \; & \; 444.7(1)        \\ 
  \hline                                                                                                                 
\de_{\QCD,\veto}^{\mu = \MZ}/\%                              \; & \; -7.0\phz        \; & \; 5.4(1)          \; & \; 2.0(1)          \; & \; -8.7(1)         \; & \; -39.2(1)        \; & \; -78.1(1)        \\ 
\de_{\QCD,\veto}^{\mathrm{var}}/\%                           \; & \; -5.6\phz        \; & \; 9.5(1)          \; & \; 13.1(1)         \; & \; 17.5(1)         \; & \; 24.8(1)         \; & \; 35.5(1)         \\ 
   \hline \hline                                                                                                         
\de_{\ga}^{\mu = \MZ}/\%                                     \; & \; 0.1\phz         \; & \; 0.1\phz         \; & \; 0.1\phz         \; & \; 0.1\phz         \; & \; 0.2\phz         \; & \; 0.4\phz         \\ 
\de_{\ga}^{\mathrm{var}}/\%                                  \; & \; 0.1\phz         \; & \; 0.1\phz         \; & \; 0.1\phz         \; & \; 0.2\phz         \; & \; 0.4\phz         \; & \; 1.0\phz         \\ 
   \hline \hline                                                                                                         
\si_{\mathrm{full, veto}}^{\mathrm{var}}/\pba                \; & \; 726.04(7)       \; & \; 276.4(1)        \; & \; 50.82(2)        \; & \; 4.260(2)        \; & \; 0.04491(3)      \; & \; 0.0003381(3)    \\ 
 \end{array} $$

\vspace{-0.5cm}
\mycaption{\label{ta:ptj_LHC7T} Integrated cross sections for different
  cuts on the $p_{\rT}$ of the leading jet at the LHC with $\sqrt{s}=8\TeV$. 
  See also caption of \refta{ta:ptj_LHC}.
  }
\end{table}
\begin{table}[p]
                                                                                                                                       $$ \begin{array}{c|rrrrrr}
                                                            \multicolumn{7}{c}{\Pp\bar\Pp \to \Pnl \Pnlb\; \mathrm{jet} + X \;\mbox{at} \;\sqrt{s} =1.96 \TeV} \\
            \hline p_{\rT,\mathrm{jet}} / \GeV & 25-\infty \;\;\; & 50-\infty \;\;\; & 75-\infty \;\;\; & 100-\infty \;\; & 200-\infty \;\; & 300-\infty \;\; \\ 
                                                                                                                                                     \hline\hline
\si_0^{\mu = \MZ}/\pba                                       \; & \; 96.613(2)       \; & \; 24.721(1)       \; & \; 8.1060(4)       \; & \; 3.0539(1)       \; & \; 0.131732(5)     \; & \; 0.0095024(6)    \\ 
\si_0^{\mathrm{var}}/\pba                                    \; & \; 93.800(2)       \; & \; 23.027(1)       \; & \; 7.1807(3)       \; & \; 2.5642(1)       \; & \; 0.090028(4)     \; & \; 0.0054067(4)    \\ 
   \hline \hline                                                                                                         
\de_{\EW}^{\mu = \MZ} / \%                                   \; & \; 0.2\phz         \; & \; 0.0\phz         \; & \; -0.6\phz        \; & \; -1.3\phz        \; & \; -4.5\phz        \; & \; -7.4\phz        \\ 
\de_{\EW}^{\mathrm{var}}/\%                                  \; & \; 0.2\phz         \; & \; 0.0\phz         \; & \; -0.5\phz        \; & \; -1.2\phz        \; & \; -4.0\phz        \; & \; -6.6\phz        \\ 
   \hline \hline                                                                                                         
\de_{\QCD}^{\mu = \MZ}/\%                                    \; & \; 9.9\phz         \; & \; 19.2(1)         \; & \; 12.9(1)         \; & \; 7.0(1)          \; & \; -16.1(1)        \; & \; -38.6(1)        \\ 
\de_{\QCD}^{\mathrm{var}}/\%                                 \; & \; 11.8\phz        \; & \; 25.1(1)         \; & \; 24.0(1)         \; & \; 23.9(1)         \; & \; 24.7(1)         \; & \; 24.9(1)         \\ 
  \hline                                                                                                                 
\de_{\QCD,\veto}^{\mu = \MZ}/\%                              \; & \; 2.0\phz         \; & \; 4.1(1)          \; & \; -2.6(1)         \; & \; -9.6(1)         \; & \; -34.4(1)        \; & \; -56.5(3)        \\ 
\de_{\QCD,\veto}^{\mathrm{var}}/\%                           \; & \; 4.3\phz         \; & \; 10.6(1)         \; & \; 8.8(1)          \; & \; 7.6(1)          \; & \; 3.9(1)          \; & \; 2.1(1)          \\ 
   \hline \hline                                                                                                         
\de_{\ga}^{\mu = \MZ}/\%                                     \; & \; 0.1\phz         \; & \; 0.1\phz         \; & \; 0.1\phz         \; & \; 0.1\phz         \; & \; 0.1\phz         \; & \; 0.1\phz         \\ 
\de_{\ga}^{\mathrm{var}}/\%                                  \; & \; 0.1\phz         \; & \; 0.1\phz         \; & \; 0.1\phz         \; & \; 0.1\phz         \; & \; 0.2\phz         \; & \; 0.2\phz         \\ 
   \hline \hline                                                                                                         
\si_{\mathrm{full, veto}}^{\mathrm{var}}/\pba                \; & \; 98.074(9)       \; & \; 25.49(1)        \; & \; 7.781(7)        \; & \; 2.732(1)        \; & \; 0.0901(1)       \; & \; 0.005175(5)     \\ 
 \end{array} $$

\vspace{-0.5cm}
\mycaption{\label{ta:ptj_Tev} Integrated cross sections for different
  cuts on the $p_{\rT}$ of the leading jet at the Tevatron. See also caption
  of 
  \refta{ta:ptj_LHC}.}
\end{table}

\begin{table}[p]
                                                                                                                                         $$ \begin{array}{c|rrrrrr}
                                                                  \multicolumn{7}{c}{\Pp\Pp \to \Pnl \Pnlb\; \mathrm{jet} + X \;\mbox{at} \;\sqrt{s} =14 \TeV} \\
                           \hline \ptmiss / \GeV & 25-\infty \;\;\; & 50-\infty \;\;\; & 100-\infty \;\; & 200-\infty \;\; & 500-\infty \;\; & 1000-\infty \; \\ 
                                                                                                                                                     \hline\hline
\si_0^{\mu = \MZ}/\pba                                       \; & \; 1627.25(3)      \; & \; 582.66(3)       \; & \; 125.669(5)      \; & \; 15.0488(5)      \; & \; 0.40302(1)      \; & \; 0.0121944(2)    \\ 
\si_0^{\mathrm{var}}/\pba                                    \; & \; 1608.99(2)      \; & \; 566.52(2)       \; & \; 116.092(4)      \; & \; 12.2928(4)      \; & \; 0.243365(6)     \; & \; 0.00523456(9)   \\ 
   \hline \hline                                                                                                         
\de_{\EW}^{\mu = \MZ} / \%                                   \; & \; 0.0\phz         \; & \; -0.3\phz        \; & \; -2.0\phz        \; & \; -6.3\phz        \; & \; -16.5\phz       \; & \; -27.4(1)        \\ 
\de_{\EW}^{\mathrm{var}}/\%                                  \; & \; 0.0\phz         \; & \; -0.2\phz        \; & \; -1.8\phz        \; & \; -5.9\phz        \; & \; -15.8\phz       \; & \; -26.2(1)        \\ 
   \hline \hline                                                                                                         
\de_{\QCD}^{\mu = \MZ}/\%                                    \; & \; 8.5\phz         \; & \; 48.1(1)         \; & \; 48.3(1)         \; & \; 35.5(1)         \; & \; -2.6(1)         \; & \; -44.2(1)        \\ 
\de_{\QCD}^{\mathrm{var}}/\%                                 \; & \; 8.4\phz         \; & \; 48.4(1)         \; & \; 51.1(1)         \; & \; 47.6(1)         \; & \; 40.4\phz        \; & \; 38.5\phz        \\ 
  \hline                                                                                                                 
\de_{\QCD,\veto}^{\mu = \MZ}/\%                              \; & \; -8.8\phz        \; & \; 18.8(1)         \; & \; 16.5(1)         \; & \; 5.6(1)          \; & \; -27.9(1)        \; & \; -64.4(1)        \\ 
\de_{\QCD,\veto}^{\mathrm{var}}/\%                           \; & \; -8.0\phz        \; & \; 20.9(1)         \; & \; 22.6(1)         \; & \; 22.8(1)         \; & \; 21.6(1)         \; & \; 24.3\phz        \\ 
   \hline \hline                                                                                                         
\de_{\ga}^{\mu = \MZ}/\%                                     \; & \; 0.0\phz         \; & \; 0.1\phz         \; & \; 0.1\phz         \; & \; 0.1\phz         \; & \; 0.1\phz         \; & \; 0.2\phz         \\ 
\de_{\ga}^{\mathrm{var}}/\%                                  \; & \; 0.0\phz         \; & \; 0.1\phz         \; & \; 0.1\phz         \; & \; 0.1\phz         \; & \; 0.2\phz         \; & \; 0.5\phz         \\ 
   \hline \hline                                                                                                         
\si_{\mathrm{full, veto}}^{\mathrm{var}}/\pba                \; & \; 1480.7(2)       \; & \; 683.9(2)        \; & \; 140.37(4)       \; & \; 14.388(4)       \; & \; 0.2581(1)       \; & \; 0.005161(3)     \\ 
 \end{array} $$

  \vspace{-0.5cm}
  \mycaption{\label{ta:ptmiss_LHC} Integrated cross sections for different
    cuts on the missing $p_{\rT}$ at the LHC with 
    $\sqrt{s}=14\TeV$. See also caption
    of 
    \refta{ta:ptj_LHC}.}
\end{table}
\begin{table}[p]
                                                                                                                                         $$ \begin{array}{c|rrrrrr}
                                                                   \multicolumn{7}{c}{\Pp\Pp \to \Pnl \Pnlb\; \mathrm{jet} + X \;\mbox{at} \;\sqrt{s} =8 \TeV} \\
                           \hline \ptmiss / \GeV & 25-\infty \;\;\; & 50-\infty \;\;\; & 100-\infty \;\; & 200-\infty \;\; & 500-\infty \;\; & 1000-\infty \; \\ 
                                                                                                                                                     \hline\hline
\si_0^{\mu = \MZ}/\pba                                       \; & \; 782.38(1)       \; & \; 263.82(1)       \; & \; 50.954(2)       \; & \; 4.9461(2)       \; & \; 0.075842(2)     \; & \; 0.00086825(2)   \\ 
\si_0^{\mathrm{var}}/\pba                                    \; & \; 768.26(1)       \; & \; 252.79(1)       \; & \; 45.631(2)       \; & \; 3.8100(1)       \; & \; 0.0409508(8)    \; & \; 0.0003054(3)    \\ 
   \hline \hline                                                                                                         
\de_{\EW}^{\mu = \MZ} / \%                                   \; & \; 0.1\phz         \; & \; -0.2\phz        \; & \; -1.8\phz        \; & \; -6.0\phz        \; & \; -15.8\phz       \; & \; -26.6(1)        \\ 
\de_{\EW}^{\mathrm{var}}/\%                                  \; & \; 0.1\phz         \; & \; -0.1\phz        \; & \; -1.6\phz        \; & \; -5.6\phz        \; & \; -15.1\phz       \; & \; -25.2(1)        \\ 
   \hline \hline                                                                                                         
\de_{\QCD}^{\mu = \MZ}/\%                                    \; & \; 8.2\phz         \; & \; 43.7(1)         \; & \; 39.9(1)         \; & \; 25.3(1)         \; & \; -12.4(1)        \; & \; -57.2(1)        \\ 
\de_{\QCD}^{\mathrm{var}}/\%                                 \; & \; 8.8\phz         \; & \; 45.8(1)         \; & \; 46.7(1)         \; & \; 44.8(1)         \; & \; 43.4\phz        \; & \; 47.1\phz        \\ 
  \hline                                                                                                                 
\de_{\QCD,\veto}^{\mu = \MZ}/\%                              \; & \; -7.0\phz        \; & \; 16.9(1)         \; & \; 10.8(1)         \; & \; -1.4(1)         \; & \; -33.5(1)        \; & \; -72.7(1)        \\ 
\de_{\QCD,\veto}^{\mathrm{var}}/\%                           \; & \; -5.6\phz        \; & \; 20.5(1)         \; & \; 20.5(1)         \; & \; 22.4(1)         \; & \; 27.3\phz        \; & \; 35.8(1)         \\ 
   \hline \hline                                                                                                         
\de_{\ga}^{\mu = \MZ}/\%                                     \; & \; 0.1\phz         \; & \; 0.1\phz         \; & \; 0.1\phz         \; & \; 0.1\phz         \; & \; 0.2\phz         \; & \; 0.4\phz         \\ 
\de_{\ga}^{\mathrm{var}}/\%                                  \; & \; 0.1\phz         \; & \; 0.1\phz         \; & \; 0.1\phz         \; & \; 0.2\phz         \; & \; 0.4\phz         \; & \; 1.0\phz         \\ 
   \hline \hline                                                                                                         
\si_{\mathrm{full, veto}}^{\mathrm{var}}/\pba                \; & \; 726.04(7)       \; & \; 304.53(9)       \; & \; 54.29(1)        \; & \; 4.457(1)        \; & \; 0.04611(1)      \; & \; 0.0003407(2)    \\ 
 \end{array} $$

  \vspace{-0.5cm}
  \mycaption{\label{ta:ptmiss_LHC8T} Integrated cross sections for different
    cuts on the missing $p_{\rT}$ at the LHC with 
    $\sqrt{s}=8\TeV$. See also caption
    of 
    \refta{ta:ptj_LHC}.}
\end{table}

\begin{table}
                                                                                                                                         $$ \begin{array}{c|rrrrrr}
                                                            \multicolumn{7}{c}{\Pp\bar\Pp \to \Pnl \Pnlb\; \mathrm{jet} + X \;\mbox{at} \;\sqrt{s} =1.96 \TeV} \\
                         \hline \ptmiss / \GeV & 25-\infty \;\;\; & 50-\infty \;\;\; & 75-\infty \;\;\; & 100-\infty \;\; & 200-\infty \;\; & 300-\infty \;\; \\ 
                                                                                                                                                     \hline\hline
\si_0^{\mu = \MZ}/\pba                                       \; & \; 96.613(2)       \; & \; 24.721(1)       \; & \; 8.1060(4)       \; & \; 3.0539(1)       \; & \; 0.131732(5)     \; & \; 0.0095024(6)    \\ 
\si_0^{\mathrm{var}}/\pba                                    \; & \; 93.800(2)       \; & \; 23.027(1)       \; & \; 7.1807(3)       \; & \; 2.5642(1)       \; & \; 0.090028(4)     \; & \; 0.0054067(4)    \\ 
   \hline \hline                                                                                                         
\de_{\EW}^{\mu = \MZ} / \%                                   \; & \; 0.2\phz         \; & \; 0.2\phz         \; & \; -0.4\phz        \; & \; -1.1\phz        \; & \; -4.2\phz        \; & \; -7.0\phz        \\ 
\de_{\EW}^{\mathrm{var}}/\%                                  \; & \; 0.2\phz         \; & \; 0.2\phz         \; & \; -0.2\phz        \; & \; -0.8\phz        \; & \; -3.5\phz        \; & \; -6.0\phz        \\ 
   \hline \hline                                                                                                         
\de_{\QCD}^{\mu = \MZ}/\%                                    \; & \; 9.9\phz         \; & \; 38.1\phz        \; & \; 31.5\phz        \; & \; 25.2\phz        \; & \; 2.1\phz         \; & \; -16.9\phz       \\ 
\de_{\QCD}^{\mathrm{var}}/\%                                 \; & \; 11.8\phz        \; & \; 43.6\phz        \; & \; 41.3\phz        \; & \; 39.6\phz        \; & \; 35.6\phz        \; & \; 34.8\phz        \\ 
  \hline                                                                                                                 
\de_{\QCD,\veto}^{\mu = \MZ}/\%                              \; & \; 2.0\phz         \; & \; 20.0\phz        \; & \; 10.7\phz        \; & \; 5.9\phz         \; & \; -12.5\phz       \; & \; -28.3\phz       \\ 
\de_{\QCD,\veto}^{\mathrm{var}}/\%                           \; & \; 4.3\phz         \; & \; 26.4\phz        \; & \; 21.7\phz        \; & \; 21.8\phz        \; & \; 23.1\phz        \; & \; 25.5\phz        \\ 
   \hline \hline                                                                                                         
\de_{\ga}^{\mu = \MZ}/\%                                     \; & \; 0.1\phz         \; & \; 0.1\phz         \; & \; 0.1\phz         \; & \; 0.1\phz         \; & \; 0.1\phz         \; & \; 0.1\phz         \\ 
\de_{\ga}^{\mathrm{var}}/\%                                  \; & \; 0.1\phz         \; & \; 0.1\phz         \; & \; 0.1\phz         \; & \; 0.1\phz         \; & \; 0.2\phz         \; & \; 0.2\phz         \\ 
   \hline \hline                                                                                                         
\si_{\mathrm{full, veto}}^{\mathrm{var}}/\pba                \; & \; 98.074(9)       \; & \; 29.178(4)       \; & \; 8.729(2)        \; & \; 3.1058(5)       \; & \; 0.10782(2)      \; & \; 0.006474(1)     \\ 
 \end{array} $$

  \vspace{-0.5cm}
  \mycaption{\label{ta:ptmiss_Tev} Integrated cross sections for different
    cuts on the missing $p_{\rT}$ at the Tevatron with 
    $\sqrt{s}=1.96\TeV$. See also caption
    of 
    \refta{ta:ptj_LHC}.}
\end{table}

In \reftas{ta:ptj_LHC}--\ref{ta:ptmiss_Tev} we summarize our 
results for LO integrated cross sections and the corresponding relative 
corrections for different cuts on the transverse momentum of the 
leading jet and the missing transverse momentum.
All other cuts and the corresponding
event selection follow our default choice as introduced in
\refse{se:cuts}.
In \reffis{fi:ptj_all}--\ref{fi:ptmiss_all} we present the 
corresponding results for differential distributions.
Both, in the tables and the figures we also show the NLO cross section
$\si_{\mathrm{full},\mathrm{veto}}^{\mathrm{var}}$ including the EW
corrections, the photon-induced processes, and the QCD 
corrections with the jet veto for the variable scale choice.
All results are discussed in detail in the
following subsections.

\subsubsection{Transverse momentum and rapidity of the leading jet}
\label{se:jet_obs}

\reftas{ta:ptj_LHC}--\ref{ta:ptj_Tev} show the predictions for different
cuts on the transverse momentum of the leading jet,
$p_{\rT,\mathrm{jet}}$, at the LHC and the Tevatron. The corresponding
differential cross sections are displayed in Figure~\ref{fi:ptj_all} for
the LHC with $\sqrt{s}=14\TeV$ and the Tevatron.  The qualitative
features of the EW corrections are similar to the results obtained
earlier for $\Plp\Plm+\mathrm{jet}$ production.

At high CM energies the generic well-known (negative) Sudakov logarithms
of the form $\ln^2(\hat{s}/M_{\PZ}^2)$ in the virtual EW corrections
lead to large corrections, e.g.\ $-25\%$ for $p_{\rT,\mathrm{jet}} \sim
1\TeV$.  If the integrated cross section is not dominated by events with
high CM energy (left columns in the tables) the EW corrections for
monojet production are negligibly small, at the permille level. As
expected, the relative EW corrections are neither particularly sensitive
to the scale choice nor to the CM energy of the LHC.  Also at the
Tevatron, the qualitative features of the corrections are very
similar. The onset of the Sudakov dominance is visible as can be seen in
\refta{ta:ptj_Tev} and \reffi{fi:ptj_all}.  However, owing to the
limited kinematic reach of the Tevatron the effects are not very
pronounced.

The dominating Sudakov logarithms lead to corrections to the
underlying process of $\PZ+\mathrm{jet}$ production and do not depend on
the specific decay channel. Hence, the similarity of the large EW
corrections presented here with the EW corrections to
$\Plp\Plm+\mathrm{jet}$ production presented in \citere{Denner:2011vu}
is not surprising. At large $p_{\rT,\mathrm{jet}}$ the corrections
differ only at the level of 1\%. Given this fact, also the agreement
with earlier on-shell calculations \citeres{Kuhn:2004em,Kuhn:2005az}
within $1{-}2\%$ observed before still holds (see Figure\ 5 in
\citere{Kuhn:2005az}). In other words, it
is well justified to use the Sudakov approximation to reliably
predict the transverse-momentum distribution of this very observable.
The residual differences at the 1\% level, which might be of interest for 
a precision determination of cross-section ratios for different 
\PZ-boson decay modes, are discussed in detail in \refse{se:ratio}.

The contribution $\delta_\gamma$ from the photon-induced processes are
tiny and do not reach the percent level even for large cut values.
This indeed justifies to safely neglect the NLO QCD 
corrections to the photon-induced channels.

Turning to the NLO QCD results at the LHC at 14 TeV, we observe results
similar to $l^+l^-+\mathrm{jet}$ production presented in
\citere{Denner:2011vu}. From the QCD point of view, the
corrections to $l^+l^-+\mathrm{jet}$ production and $\Pnl\Pnlb +
\mathrm{jet}$ production are almost equivalent. However, the acceptance
and isolation cuts necessarily differ and can lead to quantitatively
different corrections while the qualitative picture is similar.  The
corrections turn out to be accidentally small ($\sim 10\%$)
for our default cuts in the monojet
case. As discussed above, the cross section for large cut values of
$p_{\rT,\mathrm{jet}}$ contains large contributions from two-jet events
with relatively little missing transverse momentum, i.e.\ events having
little in common with typical monojet events but resulting in huge
positive corrections. The correction $\delta_\mathrm{QCD}^{\mu=\MZ}$ is
smaller than $\delta_\mathrm{QCD}^{\mathrm{var}}$, because it is defined
relative to a larger LO cross section. In absolute size, however, the
two NLO corrections are similar. Using the jet veto proposed at the end
of \refse{se:basic_cuts}, the corrections are reduced, and
$\delta_\mathrm{QCD}^{\mathrm{var}}$ only rises to the $20\%$ level
for large cut values at the 14 TeV LHC. 
As expected, at large $p_{\rT,\mathrm{jet}}$, the overestimated 
LO cross section with a fixed scale
receives large negative corrections and the discrepancy of the LO results for 
the two scale choices is largely removed by including the NLO corrections. 

At CM energy $\sqrt{s}=8\TeV$ (see \refta{ta:ptj_LHC7T}) and at the Tevatron
(see \refta{ta:ptj_Tev}), the same qualitative results are found.
Using the variable scale, increasingly negative
corrections with increasing $p_{\rT,\mathrm{jet}}$ can be avoided,
in particular when employing a jet veto, and 
stable results are obtained. At the Tevatron, the jet veto is not 
as important because of its kinematical limitations.

\bfi     
\bce
\includegraphics[width=16.1cm]{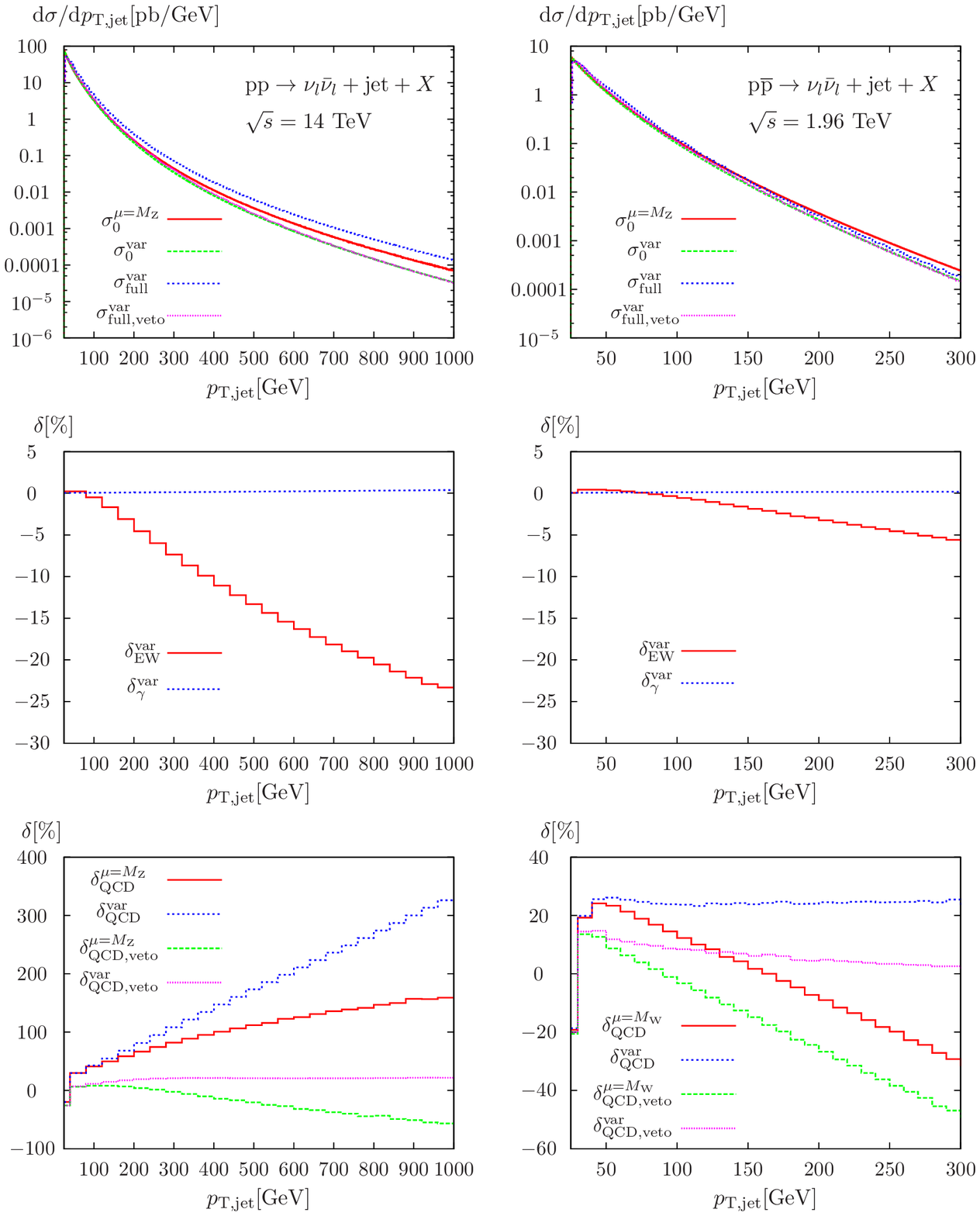} 
\ece
\mycaption{\label{fi:ptj_all} LO and fully corrected distribution (top),
  corresponding relative EW and photon-induced corrections (middle), and 
  relative QCD corrections (bottom) for the transverse momentum
  of the leading jet at the LHC (left) and the Tevatron (right). 
  }
\efi 

\bfi     
\bce
\includegraphics[width=16.1cm]{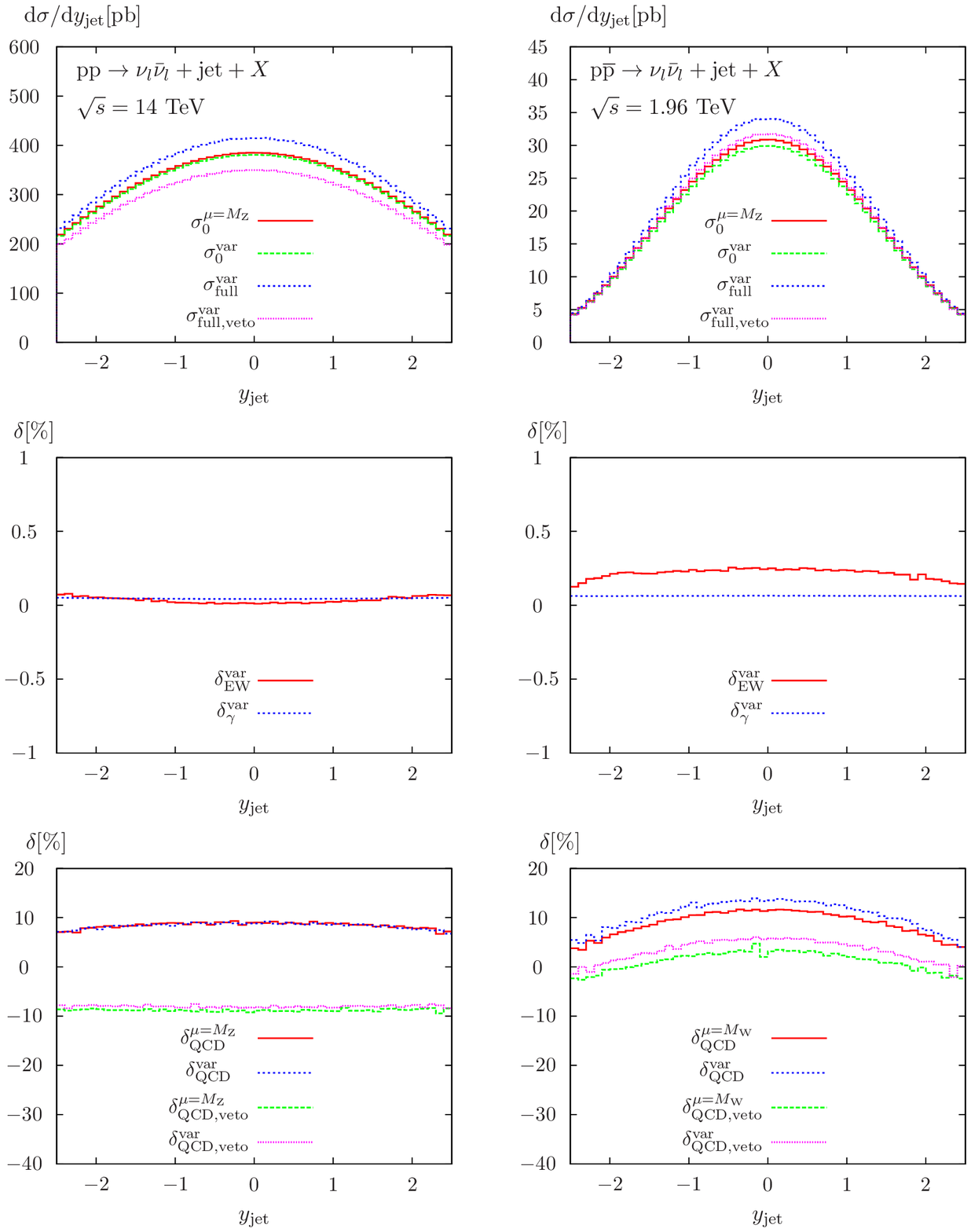} 
\ece
\mycaption{\label{fi:yj_all} LO and fully corrected distribution (top),
  corresponding relative EW and photon-induced corrections (middle), and 
  relative QCD corrections (bottom) for the rapidity
  of the leading jet at the LHC (left) and the Tevatron (right). 
  }
\efi 

Concerning the rapidity of the leading jet $y_{\mathrm{jet}}$ displayed
in \reffi{fi:yj_all}, the EW corrections both at the LHC and the
Tevatron are flat and extremely small in size, resembling the
corrections to the total cross sections. At the LHC, the QCD corrections
are positive (about 8\%) and give rise to nearly constant $K$-factors in
the whole rapidity range. Introducing a dedicated two-jet veto, as
detailed above, shifts the relative corrections to the level of $-10\%$. 
At the Tevatron, the effect of the jet veto turns out to be
smaller than at the LHC, whereas the scale dependence is somewhat
larger.

\subsubsection{Missing transverse momentum}

The missing transverse momentum, i.e.\ the transverse momentum of the
\PZ~boson, equals the leading-jet $p_{\rT}$ at LO, while at NLO the
two observables become different if an additional bremsstrahlung
particle is present. The relative EW corrections for different cut
values of $\ptmiss$ in
Tables~\ref{ta:ptmiss_LHC}--\ref{ta:ptmiss_Tev} as well as the
relative corrections to the differential distributions presented in
\reffi{fi:ptmiss_all} are completely dominated by virtual
contributions and, hence, hardly differ from the corresponding values
in Tables~\ref{ta:ptj_LHC}--\ref{ta:ptj_Tev} and \reffi{fi:ptj_all},
respectively. In contrast, the nature of the corresponding QCD
corrections changes dramatically. The huge positive corrections at
high $p_{\rT,\mathrm{jet}}$
induced by events with two hard back-to-back jets are absent, since now a
large missing momentum has to be balanced such that back-to-back jets
are kinematically suppressed.  Therefore, the two-jet veto only slightly
changes the corresponding relative corrections.
 If a two-jet veto is applied, the missing
transverse momentum has to be balanced by one hard jet, leading to
similar values of $p_{\rT,\mathrm{jet}}$ and $\ptmiss$ for
such events. Therefore, at the LHC the relative QCD corrections
are very similar for both observables at large $p_{\rT}$, since
events with two hard jets emitted in the same direction are rare.  Note
in addition that the variable-scale choice leads to almost constant
$K$-factors for QCD corrections both at the LHC and the Tevatron for the
corresponding distributions (\reffi{fi:ptmiss_all}).
\bfi \bce
\includegraphics[width=16.1cm]{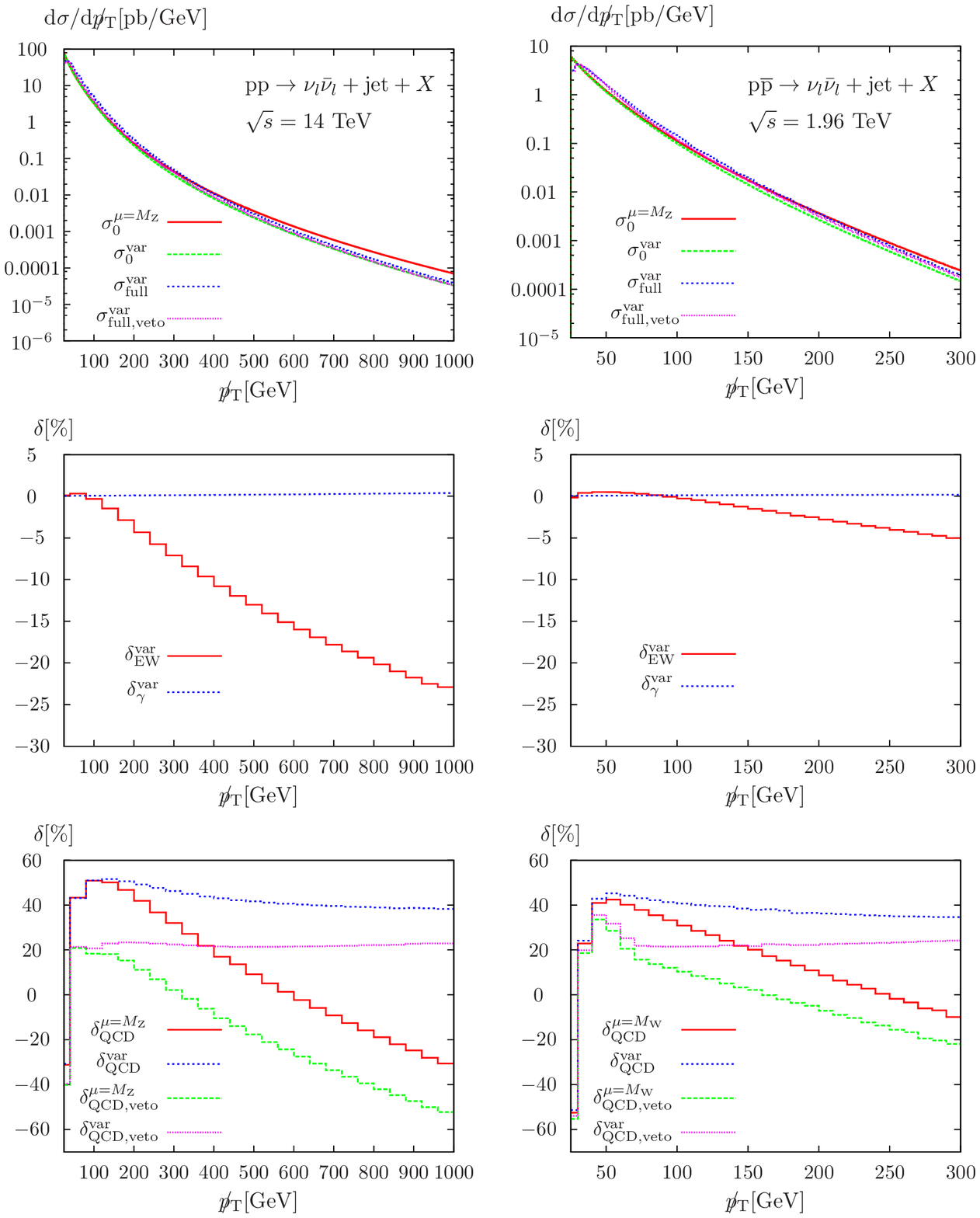} 
\ece 
\mycaption{\label{fi:ptmiss_all} LO and fully corrected distribution
  (top), corresponding relative EW and photon-induced corrections
  (middle), and relative QCD corrections (bottom) for the missing
  transverse momentum at the LHC (left) and the Tevatron (right).}
\efi

\subsubsection{Comparison of EW corrections for different leptonic
\PZ-boson decays}
\label{se:ratio}

In this section, we compare the EW corrections for 
$\PZ + \mathrm{jet}$ production with a subsequent \PZ-boson decay
into charged leptons, as investigated in detail in \citere{Denner:2011vu},
with the prediction for the monojet signature. 
As we have discussed before, due to the dominant universal 
Sudakov logarithms the EW corrections are very similar in the high-energy tails
of distributions where they matter most. However, any difference directly 
impacts the determination of the monojet cross section if a measurement of the
$l^+l^- + \mathrm{jet}$ cross section is rescaled using a 
theoretical prediction for the cross-section ratio. 

An obvious difference between the charged-lepton final state
in contrast to the neutrino final state is the presence of final-state 
radiation (FSR). FSR is often modelled in the experimental 
analysis using shower techniques so that the bulk of this difference
is assumed to be taken care of. 
Here, we want to focus on the remaining difference, which is not related to 
FSR, and matters at the 1\% level. 
Hence, in the following, we compare $\nu_l\bar\nu_l+ \mathrm{jet}$
production with $l^+l^- + \mathrm{jet}$ production
without FSR, i.e.\ we subtract all photonic (QED)
corrections to the $\PZ\to l^+l^-$ decay, which correspond to a
gauge-invariant subset of diagrams. The corresponding EW corrections are 
shown in \reffi{fi:ptmiss_comparison} for the 
distribution in the
transverse momentum $p_{\mathrm{T},V}$ of the vector boson 
and the transverse momentum $p_{\mathrm{T,jet}}$ of the leading jet
at the LHC with a CM energy of $\sqrt{s}= 8\TeV$, where $p_{\mathrm{T},V} = \ptmiss$ 
for the neutrino final state and $p_{\mathrm{T},V} = p_{\mathrm{T},l^+l^-}$ 
for the charged-lepton final state. We use the variable scale (although not
important for the relative EW corrections) and there is no visible influence of
the treatment of photon--lepton recombination for the charged-lepton final
state when FSR is subtracted.

At low transverse momenta, we observe an almost constant offset between the
corrections to the different channels. The bulk of this offset can be attributed
to the difference in the EW corrections to the partial widths of the different 
$\PZ$-boson decay modes, which are encoded in our calculation of the full
EW corrections. While the corrections to the (on-shell) $\PZ \to \nu_l\bar\nu_l$
partial width are roughly 0.9\%, the weak corrections to the $\PZ \to l^+l^-$
partial width (again subtracting the QED contribution) amount to $-0.2\%$.

Due to the universality of the Sudakov logarithms, one could expect
the 1\% offset, observed at small $p_{\mathrm{T}}$, to be constant 
over the whole $p_{\mathrm{T}}$-range. However, the EW corrections
for the neutrino final state rise faster and are slightly larger 
in absolute size than their charged-lepton counterpart at 
$p_{\mathrm{T}} = 1\TeV$. 
This difference can be attributed to the photon-exchange contribution
for the charged-lepton final state, which is part of the off-shell effects. 
For the EW correction in \reffi{fi:ptmiss_comparison}, the event definition 
(see \citere{Denner:2011vu})
in this channel only asks for a dilepton invariant mass which is bigger than 
$50\GeV$. In this case, the photon contribution is close to 10\% for 
$p_{\mathrm{T},V} = 1\TeV$ (interference 
contributions are small). Since the Sudakov logarithms in the diagrams
with an intermediate photon are much smaller 
(see Fig. 4 of \citere{Kuhn:2005gv}), the corrections for the
charged-lepton final state are indeed smaller. For a tighter selection cut, 
asking for a dilepton mass close to the \PZ\ pole, 
the photon contribution is suppressed and the difference
between the corrections in \reffi{fi:ptmiss_comparison}
is indeed constant and almost completely due to the different 
corrections to the partial width, as shown in 
\reffi{fi:ptmiss_comparison_only_Z}. This dependence of the EW corrections 
in the Sudakov regime on the setup has to be taken into account at the
$1{-}2\%$ level, when using a measurement of the charged-lepton final state to 
predict the cross section for the monojet signature.

\bfi \bce
\includegraphics[width=7.8cm]{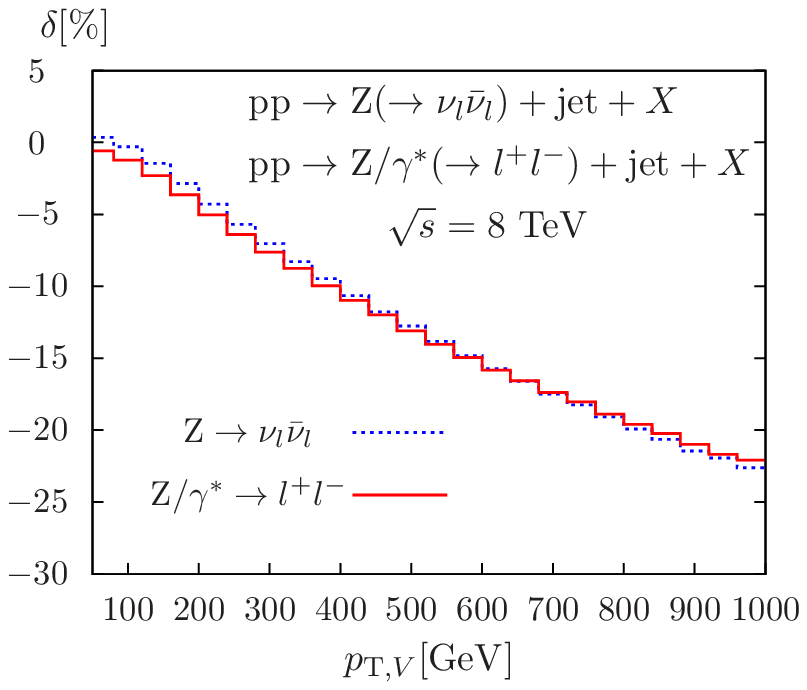} 
\includegraphics[width=7.8cm]{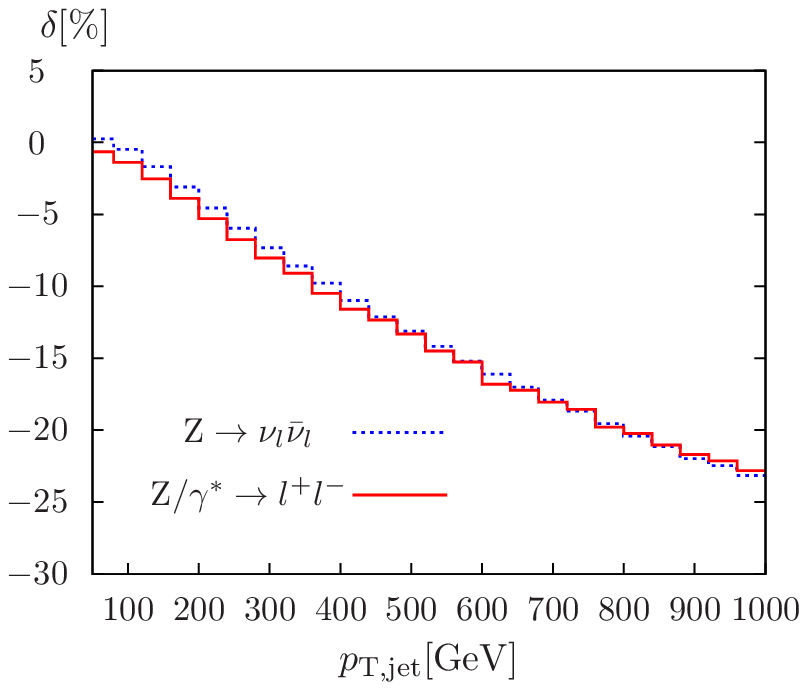} 
\ece 
\mycaption{\label{fi:ptmiss_comparison} EW corrections 
(without final-state radiation) to 
the transverse-momentum distribution of the vector boson (left) 
and the leading jet (right) for
the two different leptonic \PZ-boson decay modes, 
$\PZ \to \nu_l\bar\nu_l$ and $\PZ/\gamma^* \to l^+l^-$, in $\PZ + \mathrm{jet}$ 
production at the LHC ($\sqrt{s} = 8\TeV$).}
\efi

\bfi \bce
\includegraphics[width=7.8cm]{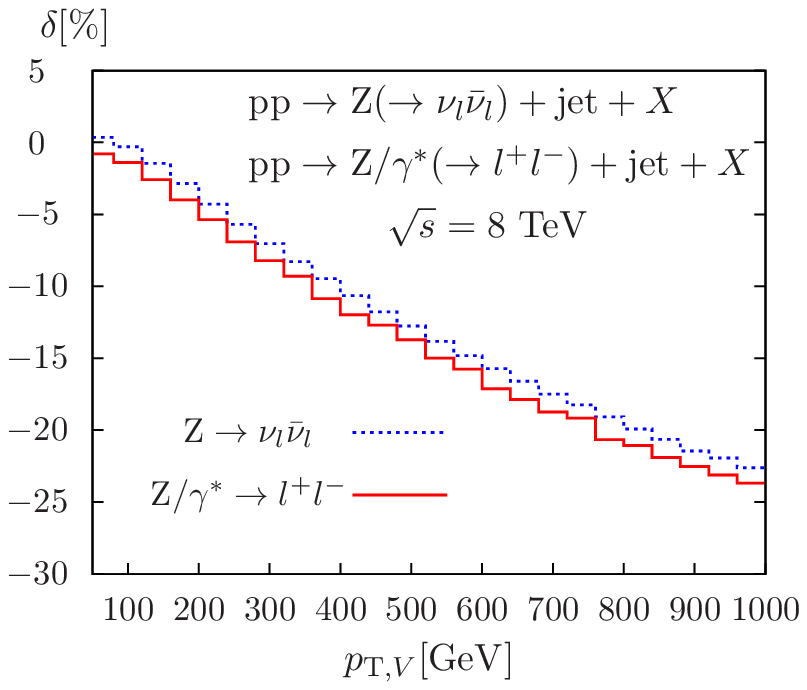} 
\ece 
\mycaption{\label{fi:ptmiss_comparison_only_Z} EW corrections 
(without final-state radiation) to 
the transverse-momentum distribution of the vector boson for
the two different leptonic \PZ-boson decay modes, 
$\PZ \to \nu_l\bar\nu_l$ and $\PZ/\gamma^* \to l^+l^-$, in $\PZ + \mathrm{jet}$ 
production at the LHC ($\sqrt{s} = 8\TeV$). In contrast to the standard setup,
a dilepton invariant-mass cut $86\GeV<M_{ll}<96\GeV$
is applied for the charged-lepton final state in order to suppress the
$\gamma^*$ component of the differential cross section.}
\efi

\section{Combination of QCD and EW effects at NLO and beyond}
\label{se:EWQCD}

In the discussion of our results in \refse{se:CSresults}, 
we combined NLO QCD and EW corrections in a purely additive 
manner to present a full NLO prediction. Mixed 
EW$\times$QCD corrections which are not part of the calculation
are not addressed at all. Hence, the naive product of 
the NLO EW and QCD corrections represents 
an error estimate for higher-order corrections including EW effects.

This limitation is particularly important in the presence of
improvements for the predictions on the QCD side: Parton-shower matching
as well as dedicated resummations have become available, and even the
calculation of next-to-next-to-leading-order QCD corrections might be
within reach~\cite{GehrmannDeRidder:2012ja}.  The increase in
theoretical accuracy as well as the decreasing experimental errors at
the LHC raise the question how to properly combine radiative EW
corrections with the best available QCD prediction.  In practice, a
similar problem arises when EW corrections are to be included in
``standard'' QCD Monte Carlo tools.

Obviously, a completely satisfactory answer to this question 
can be found in the computation
of the combined EW$\times$QCD corrections, a very difficult task
requiring involved two-loop calculations, which are beyond present
computational possibilities.
On the other hand, factorization of EW and QCD corrections is
expected to be a good approximation.
In fact, the large EW Sudakov logarithms, which completely dominate
the EW corrections where they are most relevant, are part of the hard
underlying production process. Therefore, it does not seem to be far
fetched that there is not much interplay with soft and/or collinear
physics dominating the QCD corrections.

The factorized ansatz for the EW$\times$QCD corrections can, however,
break down in certain kinematical situations and should, of course, not
be applied blindly. A good example for such a breakdown is provided by
the $p_{\rT,\mathrm{jet}}$ distribution in $V +\mathrm{jet}$ production.
In \refse{se:CSresults}, we have discussed that the tail of this
distribution is completely dominated by back-to-back jets and does not
contribute a generic NLO QCD correction to the LO kinematics. Applying
the calculated EW corrections in a factorized form does not make any
sense in this case. In fact, without a sensible jet veto, this work does
not even supply a sensible estimate for the EW corrections for this
observable at all. Only the calculation of the EW corrections to
$V+2\,\mathrm{jets}$ could improve the situation.  Nevertheless, a
dedicated jet veto solves the problem, and a factorized ansatz becomes a
good approximation.

For other distributions, like the missing transverse momentum, 
the situation is fortunately simpler. The corresponding QCD corrections
are uniform and moderate in size, as can be seen in
Fig.~\ref{fi:ptmiss_all}, and one may safely assume a factorization of
QCD and EW correction in this case, independent of a possible jet veto. 
Therefore, the best motivated 
prediction including EW corrections to
the spectrum of the missing transverse momentum is given by
\begin{equation}
  \frac{\rd\sigma^{\mathrm{best}}_{\QCD\times\EW}}{\rd \ptmiss} =
  \left[1+\delta_{\EW}(\ptmiss)\right]\frac{\rd\sigma^{\mathrm{best}}_{\QCD}}{\rd\ptmiss}\,, 
\end{equation} 
with $\delta_{\EW}(\ptmiss)$ taken from Fig.~\ref{fi:ptmiss_all}, where
the best prediction available for the QCD-corrected cross section should be
used. The residual uncertainties due to EW effects can then be estimated
by the square of the relative EW corrections or equivalently by the
two-loop EW high-energy logarithms~\cite{Kuhn:2005az}. A further reduction
of EW uncertainties would require knowledge of the sub-leading higher-order
logarithms, where potential cancellations between different logarithmic
orders are expected.

The same reasoning also holds for the distributions in $l^+l^- +
\mathrm{jet}$ production discussed in \citere{Denner:2011vu} or $\nu_l l^+ +
\mathrm{jet}$ production discussed in \citere{Denner:2009gj}. Again, the 
calculated EW corrections to the $p_{\rT,\mathrm{jet}}$ distribution are only useful
if an adequate jet veto is applied.
 
\section{Conclusions}
\label{se:concl}

Following our study on $\Plm\Plp+\mathrm{jet}$
production~\cite{Denner:2011vu}, we have presented the first
calculation of the full NLO electroweak corrections to the
production of one isolated hard jet at hadron colliders in the SM,
which is an important signal process for various new-physics models.
For all relevant observables the cross section is dominated by
on-shell $\PZ+\mathrm{jet}$ production with a subsequent leptonic
\PZ-boson decay. However, in our calculation all off-shell effects are
taken into account.

We have implemented our results in a flexible Monte Carlo code which
can model the experimental event definition at the NLO parton level.
The separation of single-jet and single-photon
production is consistently implemented by making use of the measured
quark-to-photon fragmentation function.  We have also recalculated the
NLO QCD corrections supporting a phase-space-dependent scale choice.
Photon-induced processes are included at leading order, but turn out to
be phenomenologically unimportant.

The presented electroweak corrections to the total cross sections
are at the permille level and are therefore negligible. Only in the
tails of distributions, which are dominated by large
centre-of-mass energies, the well-known Sudakov logarithms become
dominant, and the electroweak corrections increase up to $-25\%$ at 
transverse momenta of $\sim 1\TeV$.  For the $p_{\rT,\mathrm{jet}}$ 
and $\ptmiss$ distributions the
results at large transverse momenta are in good agreement with earlier
results for the $l^+l^-+\mathrm{jet}$ final state~\cite{Denner:2011vu}
as well as results obtained in the on-shell approximation for the \PZ\
boson~\cite{Kuhn:2005az}.

The QCD corrections are
 moderate for observables that are
dominated by transverse momenta below about $100\GeV$. However, they can
become extremely large (hundreds of percent) at jet transverse momenta
$p_{\rT,\mathrm{jet}}$ of some $100\GeV$ unless a sensible veto on a
second hard jet is applied.  For the $p_{\rT,\mathrm{jet}}$
distribution, we have discussed that such a jet veto is essential for
the applicability of the presented EW corrections.  In contrast, the
$\ptmiss$ distribution is quite stable against QCD corrections.
Introducing a dynamical scale flattens the $K$-factor in the high-energy
tails of the transverse-momentum distributions.

\subsection*{Acknowledgements}

This work is supported in part by the Gottfried Wilhelm Leibniz
programme of the Deutsche Forschungsgemeinschaft (DFG) and by the DFG
Sonderforschungsbereich/Transregio 9 ``Computergest\"utzte
Theoretische Teilchenphysik''.  \nopagebreak

\nopagebreak

\end{document}